\documentclass[]{JHEP3} 

\JHEPspecialurl{http://jhep.sissa.it/JOURNAL/JHEP3.tar.gz}

\usepackage{epsfig,multicol,bbm}
\usepackage{bm}
\usepackage{amsmath}

\newcommand\fverb{\setbox\fverbbox=\hbox\bgroup\verb}
\newcommand\fverbdo{\egroup\medskip\noindent%
			\fbox{\unhbox\fverbbox}\ }
\newcommand\fverbit{\egroup\item[\fbox{\unhbox\fverbbox}]}
\newbox\fverbbox

\renewcommand{\Re}{{\rm Re \,}}
\renewcommand{\Im}{{\rm Im \,}}
\newcommand{\tr}{{\rm tr \,}}
\newcommand{\Tr}{{\rm Tr}}

\newcommand{\Dirac}{{\bf D}}

\newcommand{\thru}[1]{ \mathrel{\mathop{#1\!\!\!\!/}}}
\newcommand{\thrur}[1]{\mathrel{\mathop{#1\!\!\!/}}}
\newcommand{\thrul}[1]{\mathrel{\mathop{#1\!\!\!\!\!/}}}

\newcommand{\D}{D}

\newcommand{\sm}{\mbox{{\textit{m}}}}

\newcommand{\kCP}{\kappa_{{\rm CP}}}

\newcommand{\llangle}{\left\langle}
\newcommand{\rrangle}{\right\rangle}

\newcommand{\mlr}{m_{LR}}
\newcommand{\mrl}{m_{RL}}

\newcommand{\w}{{w}}

\newcommand{\comments}[1]{} 

\title{CP violation in the effective action of the Standard Model}

\author{Carmen Garc{\'\i}a-Recio and Lorenzo Luis Salcedo\\
Departamento de F{\'\i}sica At\'omica, Molecular y
Nuclear, Universidad de Granada, E-18071 Granada, Spain\\
	E-mail: \email{g\_recio@ugr.es}, \email{salcedo@ugr.es}}

\received{\today} 		
\accepted{\today}		

\preprint{}	

\abstract{ Following a suggestion by Smit, the CP odd terms of
  the effective action of the Standard Model, obtained by integration
  of quarks and leptons, are computed to sixth order within a strict
  covariant derivative expansion approach. No other approximations are
  made. The final result so derived includes all Standard Model gauge
  fields and Higgs. Remarkably, at the order considered in this work,
  all parity violating contributions turn out to be zero. Non vanishing
  CP violating terms are obtained in the C-odd P-even sector. These
  are several orders of magnitude larger than perturbative estimates.
  Various unexpected regularities in the final result are noted.  }

\keywords{Nonperturbative Effects, Standard Model, CP violation,
Baryogenesis}

\begin{document} 



\newpage

\section{Introduction}

Whereas CPT symmetry is preserved of necessity in any theory with
Lorentz invariance, locality and unitarity \cite{Streater:1989vi} no
such mechanism is at work to preserve CP \cite{Christenson:1964fg}.
Nevertheless, CP is very weakly broken in the Standard Model
\cite{Beneke:2002fz,Nir:1998pg}. No breaking is detected in the
strong sector, where the coupling constant of the operator $G_{\mu\nu}
\tilde{G}^{\mu\nu}$ is compatible with zero \cite{Peccei:1998jt}. In
the electroweak sector no breaking is possible for less than three
generations and even in this case the breaking would not occur in the
presence of mass degeneration, as is the case in the lepton sector for
massless neutrinos \cite{Kobayashi:1973fv}. So there is CP violation
in the Standard Model but it is rather small, as compared to the
maximal breaking of C or P in the electroweak sector.

The amount of CP breaking is relevant in early universe baryogenesis
\cite{Cohen:1993nk}. Baryon asymmetry generation is assumed to take
place near the electroweak transition temperature, $T$ of the order of
$100\,\text{GeV}$. At such high temperatures quark masses, except that
of the top quark, can be considered small. This suggests to treat them
as a perturbation.  Since these masses follow from the Yukawa coupling
of quarks to the Higgs field, this is equivalent to treat those Yukawa
vertices perturbatively. However, as noted above, CP violation is
elusive as no CP breaking term can be produced at low orders. The
simplest such term is the Jarlskog determinant which appears at order
twelve \cite{Jarlskog:1985ht}
\begin{equation}
\Delta = J(m_u^2-m_c^2)(m_c^2-m_t^2)(m_t^2-m_u^2)
(m_d^2-m_s^2)(m_s^2-m_b^2)(m_b^2-m_d^2) \,.
\label{eq:1.1}
\end{equation}
where $J$ is the dimensionless Jarlskog invariant, constructed with
the Cabibbo-Kobayashi Maskawa matrix. This term has dimension twelve
and by dimensional counting it is usually assumed to enter in the
effective action scaled by $T^{12}$. At the electroweak transition
temperature the ratio $\Delta/T^{12}$ is extremely small, of the
order of $10^{-19}$. If this estimate is correct this poses a problem
to account for baryogenesis using the Standard Model \cite{Rubakov:1996vz}.

A simpleminded transcription of the above estimate to the zero temperature
case can be achieved by simply replacing the scale $T$ by the Higgs field
condensate $v=246\,\text{GeV}$, $\Delta/v^{12}= 10^{-24}$. This is
several orders of magnitude smaller than CP violation as measured in
meson decays, where dimensionless parameters are of the order of
$10^{-3}$ \cite{Amsler:2008zzb}. However, the direct use of the
Jarlskog determinant is not justified at zero temperature, where the
quark masses can no longer be treated perturbatively.

Smit \cite{Smit:2004kh} made the observation that a non a perturbative
treatment would yield in a natural way much larger couplings for CP
breaking operators, as such couplings would come out as rational
functions (with logarithms) of the quark masses times the Jarlskog
invariant.  Specifically it was proposed to study the effective action
of the Standard Model obtained after integration of the fermions in
the theory. The full functional is, of course, beyond an exact
computation and some type of classification and selection of the
resulting terms is required.  The proposal was then to organize the
terms within a covariant derivative expansion, which being non
perturbative has the potential of yielding a more reliable estimate for
the couplings.

The study of the leading order terms in the abnormal parity sector was
undertaken in \cite{Smit:2004kh}. This is the sector driven by the
Levi-Civita pseudo-tensor and includes the Wess-Zumino-Witten term. In
the Standard Model this is equivalent to the parity odd sector.  The
corresponding operators have dimension 4, counting only the dimension
carried by the dynamical fields except the Higgs.  Unfortunately no
non vanishing contribution was found to fourth order, although CP
breaking contributions were expected at dimension 6
\cite{Smit:2004kh}. Such dimension 6 operators, with non vanishing
coupling, have been found in \cite{Hernandez:2008db} where also the
abnormal parity sector was studied. As expected, the CP violating term
found is indeed sizable as compared to perturbative estimates.

The two calculations of CP violating terms in the Standard Model
effective action just described are based on the technique introduced
in \cite{Salcedo:2000hp,Salcedo:2000hx}. In this approach the full
effective action for a generic theory of fermions coupled to chiral
gauge fields is computed within a strict derivative expansion, to a
given order.  \cite{Smit:2004kh} uses directly the result in
\cite{Salcedo:2000hx} which holds to fourth order, and particularizes
it to the Standard Model while selecting just the CP breaking terms.
\cite{Hernandez:2008db} carries out the same reduction from general to
particular after extending the generic calculation to sixth order.
This is done using the worldline formalism to deal with Dirac traces
and momentum integration, as explained in \cite{Hernandez:2007ng}, as
an alternative to do the same thing with the techniques applied in
\cite{Salcedo:2000hx}.

In the present work we also undertake the calculation of the CP odd
component of the Standard Model effective action at zero temperature
derived by integration of quarks and leptons, and organize the terms
so obtained by means of a covariant derivative expansion. However,
unlike previous calculations, ours is carried out from scratch by
applying the recently derived technique described in
\cite{Salcedo:2008tc,Salcedo:2008bs}. The difference with previous
approaches is that we particularize very early our treatment to the
Standard Model and this allows us to select from the beginning terms
which are candidate to break CP invariance and neglect irrelevant CP
even terms. This is useful as CP breaking imposes very restrictive
conditions and selects very few candidates. Another difference is that
we consider terms of normal parity (i.e., P even and consequently C
odd) as well as of abnormal parity (P odd, C even).  Seemingly, in the
literature, the CP breaking terms have been assumed to be of abnormal
parity only.  Perhaps this is because in the so called strong CP
problem, the CP breaking terms involve the Levi-Civita pseudo-tensor
in the gluon sector or $\gamma_5$ in the quark sector
\cite{Jackiw:1976pf,Callan:1976je}. Moreover, the simplest CP odd
terms one can write for the effective action are also odd under
intrinsic parity, e.g., $\tr(F_{\mu\nu}\tilde{F}^{\mu\nu})$
\cite{Shaposhnikov:1987pf}, where $F_{\mu\nu}$ is constructed with the
covariant derivative of the electroweak group
$\text{SU}(2)\times\text{U}(1)$. In addition, the CP odd and P odd
sector is of interest in the study of electric dipole moment.
Nevertheless, it is perfectly possible to write down operators,
constructed with the Standard Model gauge fields and the Higgs, having
normal intrinsic parity and odd under CP transformations.

We carry out a detailed calculation including all Standard Model
fields and all CP violating contributions to order six in the
covariant derivative expansion. Diagrammatically, this consists of all
one-loop Feynman graphs with fermions running on the loop and up to
six gauge fields or derivatives (four-momenta) attached as external
legs and any number of Higgs fields coupled to the quark or leptons.
The result is given in the unitary gauge. The calculation presented is
largely self-contained and some of the main conclusions obtained can
be reached by a by-hand computation. This is the case for our most
unexpected finding, namely, that there are CP violating terms of order
six in the normal parity sector of the Standard Model, but all terms
vanish in the abnormal parity sector. This is at variance with
the result in \cite{Hernandez:2008db}. It is not clear to us from
where the discrepancy arises as the two calculations are conceptually
similar although technically different. In any case we have
double-checked our results to confirm this conclusion.

The paper is organized as follows. Section \ref{sec:2} summarizes a
number of definitions, formulas and techniques relative to generic
chiral gauge theories which will then be applied to the Standard
Model. In section \ref{sec:3} we cast the fermionic sector of the
Standard Model in the format previously described for generic chiral
gauge theories. In section \ref{sec:4} features of CP violation at the
level of the effective action are discussed. In section \ref{sec:5}
the relevant momentum integrals which appear in the calculation, as
well as the associated selection rules, are obtained. Section
\ref{sec:6} discusses the covariant derivative expansion within our
approach to the effective action of the Standard Model. In section
\ref{sec:7} the chiral invariant approach devised for generic theories
is applied to the Standard Model in a way that allows to easily remove
irrelevant CP even terms.  Section \ref{sec:8} reproduces the previous
result in \cite{Smit:2004kh} verifying that there are no CP breaking
terms driven by operators of order four in the abnormal parity sector,
and this result is extended to the normal parity sector as well. In
section \ref{sec:9} we present a preliminary calculation for the
particular case where Higgs field derivatives are neglected. This is
of interest as this covers the result obtained in
\cite{Hernandez:2008db}. The cancellation of the abnormal parity
contribution is made manifest there within a transparent calculation.
Section \ref{sec:10} presents the full result of our computation,
Eq.~(\ref{eq:10.4}).  Various surprising regularities in the result
are noted.  The form of the loop function controlling the CP breaking
operator is discussed in section \ref{sec:12} and how it is affected
by infrared enhancement in the physically relevant chiral limit.  Our
conclusions are summarized in section \ref{sec:11}.

\section{Chiral gauge fermions}
\label{sec:2}

In this section we collect important practical results relative to
generic chiral gauge fermion theories which will be applied
subsequently to the Standard Model.

\subsection{The effective action}

The Lagrangian describing the coupling of spin 1/2 fermions ($\psi$) to chiral
gauge fields ($V_{L,R}$) and spin zero fields ($m_{LR}$, $m_{RL}$) can
be cast in the general form
\begin{equation}
{\cal L}(x)=\bar{\psi}(x)\,i\Dirac(x) \, \psi(x) \,,
\end{equation}
where\footnote{We use Minkowskian signature $(+---)$ and
$\{\gamma^\mu,\gamma^\nu\}=2g^{\mu\nu}$,
$\gamma_5=+i\gamma^0\gamma^1\gamma^2\gamma^3$, $\epsilon_{0123}=+1$.}
\begin{equation}
i\Dirac(x) =
(i\thrur{\partial} - \thrul{V}_R\!(x) )P_R
+(i\thrur{\partial} - \thrul{V}_L\!(x) )P_L
-\mlr(x) P_R -\mrl(x) P_L ,
\end{equation}
and
\begin{eqnarray}
P_R &=& \frac{1}{2}(1+\gamma_5),\quad P_L=\frac{1}{2}(1-\gamma_5),
\end{eqnarray}
and the external fields $\mlr$, $\mrl$, $V_{R,L}$ are matrices in some
internal space.  Unitarity requires
\begin{eqnarray}
\mlr^\dagger(x)=\mrl(x),\quad
V_{R,\mu}^\dagger(x)=V_{R,\mu}(x),\quad
V_{L,\mu}^\dagger(x)=V_{L,\mu}(x) .
\end{eqnarray}

It will prove convenient to write the Lagrangian in matricial form,
namely,
\begin{equation}
{\cal L}(x) = (\bar{\psi}_L,\bar{\psi}_R) \left(\begin{matrix}
 -\mlr & i\thrur{\partial} - \thrul{V}_L\!(x)  \\
i\thrur{\partial} - \thrul{V}_R\!(x) & -\mrl
\end{matrix}
\right)
 \left(\begin{matrix}
 \psi_R  \\ \psi_L
\end{matrix}
\right) .
\label{eq:5}
\end{equation}
This form is useful to expose the action of  chiral transformations.

The integration of the fermions provides the effective action
\begin{eqnarray}
Z= e^{i\Gamma} = 
\int D\bar{\psi}D\psi\,e^{i\int d^4x\,\bar{\psi}(x)\,i\Dirac \, \psi(x)},
\end{eqnarray}
which (modulo UV ambiguities) is given by 
\begin{equation}
i\Gamma[\mlr,\mrl,V_R,V_L]= \Tr\log i\Dirac .
\end{equation}

The effective action can be decomposed into a normal parity component,
$\Gamma^+$ (without Levi-Civita pseudo-tensor) and an abnormal parity
component $\Gamma^-$ (with the Levi-Civita pseudo-tensor):
\begin{eqnarray}
\Gamma =  \Gamma^+ + \Gamma^- .
\label{eq:2.8}
\end{eqnarray}
$\Gamma^\pm$ are also even/odd, respectively, under the pseudo-parity
transformation, which can be defined as the exchange of the labels
$LR$ in the external fields.

Among other symmetries, the effective action is invariant under the
transformation
\begin{eqnarray}
&&
\mlr(x) \to \mlr^*(\tilde{x}) , \quad
\mrl(x) \to \mrl^*(\tilde{x}) , \quad
\nonumber \\
&&
V_{R,\mu}(x) \to -\pi_\mu{}^\nu V_{R,\nu}^* (\tilde{x}) , \quad
V_{L,\mu}(x) \to -\pi_\mu{}^\nu V_{L,\nu}^* (\tilde{x}) , \quad
\end{eqnarray}
where $\pi^\mu{}_\nu = \text{diag}(1,-1,-1,-1)$ and $\tilde{x}=
(x^0,-\vec{x}) = \pi x$. This represents a CP transformation which we
shall call {\em full CP transformation} to distinguish it from the
physical one (which only acts on dynamical fields and not on
parameters of the Lagrangian).\footnote{Likewise, {\em full parity} is
  also a symmetry of the effective action. It consists of exchanging
  $LR$ labels and simultaneously $x\to\tilde{x}$. Pseudo-parity, which
  can be defined as any of these two transformations without the
  other, is not a symmetry for general background field
  configurations.}  The important property of this transformation is
that it does not mix different chiral sectors.

\subsection{Euclidean space}
\label{subsec:2.B}

For convenience we shall work in Euclidean space, reverting to
Minkowskian space at the end.

The Euclidean metric is $\delta_{\mu\nu}$, so we put Euclidean indices
as subindices.  The pass from Minkowskian to Euclidean variables is
achieved by the replacements\footnote{This corresponds to $\eta_4=1$
  in Ref.~\cite{Salcedo:2000hx}.}
\begin{eqnarray}
&&
(x^0,x^i) \to (-ix^0,x^i),
\nonumber
\\ 
&&
\psi(x)\to \psi(x),
\quad
\bar{\psi}(x)\to \bar{\psi}(x),
\nonumber
\\
&&
\mlr(x)\to \mlr(x),
\quad
\mrl(x)\to \mrl(x),
\nonumber
\\
&&
V_{R,0}(x)\to V_{R,0}(x),
\quad
V_{L,0}(x)\to V_{L,0}(x),
\nonumber
\\
&&
V_{R,i}(x)\to -i V_{R,i}(x), 
\quad
V_{L,i}(x)\to -i V_{L,i}(x),
\nonumber
\\
&&
\gamma^0 \to \gamma_0,
\quad
\gamma^i \to i\gamma_i,
\quad
\gamma_5 \to \gamma_5=\gamma_0\gamma_1\gamma_2\gamma_3 \, .
\end{eqnarray}
These replacements imply ${\cal L}(x)\to -{\cal L}(x)$ and so
\begin{equation}
e^{i\Gamma}
= 
\int D\bar{\psi}D\psi\,e^{i\int d^4x\,{\cal L}}
\to 
e^{-\Gamma}=e^{-\int d^4x{\cal L}(x)}
\end{equation}
with
\begin{equation}
{\cal L}(x)= \bar{\psi}(x)\Dirac \, \psi(x) \,,
\label{eq:4.3}
\end{equation}
and
\begin{equation}
\Dirac =
(\thrur{\partial} + \thrul{V}_R\!(x) )P_R
+(\thrur{\partial} + \thrul{V}_L\!(x) )P_L
+\mlr(x) P_R +\mrl(x) P_L .
\end{equation}
Also, in Euclidean space
\begin{equation}
\Gamma[\mlr,\mrl,V_R,V_L]= \Tr\log\Dirac
\,.
\label{eq:2.28}
\end{equation}

With the above prescriptions, in Euclidean space unitarity becomes
\begin{eqnarray}
\mlr^\dagger(x)=\mrl(x),\quad
V_{R,\mu}^\dagger(x)=-V_{R,\mu}(x),\quad
V_{L,\mu}^\dagger(x)=-V_{L,\mu}(x) ,
\end{eqnarray}
while the full CP transformation becomes
\begin{equation}
\mlr(x) \to \mlr^*(\tilde{x}) , \quad
\mrl(x) \to \mrl^*(\tilde{x}) , \quad
V_{R,\mu}(x) \to \pi_{\mu\nu} V_{R,\nu}^* (\tilde{x}) , \quad
V_{L,\mu}(x) \to \pi_{\mu\nu} V_{L,\nu}^* (\tilde{x}) , \quad
\label{eq:2.30}
\end{equation}
with $\pi_{\mu\nu} = \text{diag}(1,-1,-1,-1)$ and $\tilde{x}=
(x^0,-\vec{x}) = \pi x$.

An important property of the effective action in Euclidean space is
that the normal parity component, $\Gamma^+$, is real, and the
abnormal parity component, $\Gamma^-$, is purely imaginary
\cite{Alvarez-Gaume:1983ig}. This property is a consequence of
unitarity and holds at the non perturbative level.

\subsection{Chiral invariant approach to the effective action}
\label{sec:2.D}

The (Euclidean) Lagrangian (\ref{eq:4.3}) is
invariant under local chiral transformations:
\begin{equation}
\Dirac \longrightarrow \Dirac^\Omega = 
\left(
 \begin{matrix}
\Omega_L^{-1}  &  0\cr
 0   & \Omega_R^{-1} 
\end{matrix}
\right)
\left(
 \begin{matrix}
\mlr  &  \thru{\D}_L \cr
\thru{\D}_R   &   \mrl 
\end{matrix}
\right)
\left(
 \begin{matrix}
\Omega_R  &  0\cr
 0   & \Omega_L
\end{matrix}
\right) .
\label{eq:2a.19}
\end{equation}
However, as is well known, the corresponding effective action
$\Gamma[\mlr,\mrl,V_R,V_L]$ displays an anomalous variation under
chiral transformations \cite{Adler:1969gk,Bell:1969ts,Bardeen:1969md}.
The anomaly has a universal, geometrical, form and is saturated by the
gauged Wess-Zumino-Witten (WZW) term,
$\Gamma_{\text{gWZW}}[\mlr,\mrl,V_R,V_L]$, which also has a known
geometrical form \cite{Wess:1971yu,Witten:1983tw}. This means that if
the WZW term is subtracted from the effective action, the remainder,
$\Gamma_c$, is chiral invariant:
\begin{equation}
\Gamma = \Gamma_{\text{gWZW}} + \Gamma_c 
.
\end{equation}

Let us note that $\Gamma_{\text{gWZW}}$ contributes only to the
abnormal parity component. Therefore, $\Gamma^+=\Gamma_c^+$.

Recently, it has been shown that the remainder $\Gamma_c$ can be
expressed in a convenient form in which chiral invariance is manifest
\cite{Salcedo:2008tc}. Indeed, let
\begin{equation}
K
=
\mlr \mrl -  \thru{D}_L \mrl^{-1} \thru{D}_R \mrl,
\label{eq:2.20}
\end{equation}
where $D_{L,\mu}= \partial_\mu + V_{L,\mu}$ and $D_{R,\mu}=
\partial_\mu + V_{R,\mu}$, and so $K$ is a second order differential
operator. (Note that $\thru{D}_L \mrl^{-1} \thru{D}_R \mrl$ stands for
the product of four consecutive operators.) Then\footnote{In the
  notation of \cite{Salcedo:2008tc}, $K$ here is $K_L$ there, and we
  have used that $K_R$ and $K_L^\dagger$ are related by a similarity
  transformation.}
\begin{eqnarray}
\Gamma^+
&=&
-\frac{1}{2}\Re \Tr(\log K)
,
\nonumber \\
\Gamma^-
&=&
-\frac{1}{2}i\,\Im \Tr(\gamma_5\log K)
+\Gamma_{\text{gWZW}}
.
\label{eq:2.21}
\end{eqnarray}

\subsection{The derivative expansion}
\label{subsec:2.D}

The functional trace in (\ref{eq:2.28}) can not be computed in closed
form in general. This suggests to use instead some systematic
expansion to address its determination. In the derivative expansion
scheme the effective action contributions are classified according to
the number of covariant derivatives they carry. In this counting the
spin zero fields, $\mlr$, $\mrl$, count as order zero whereas the
derivatives $\partial_\mu$ or gauge fields, $V_{L,\mu}$, $V_{R,\mu}$,
count as first order. Technically, this means to consider the family
of Dirac operators
\begin{equation}
\Dirac_{t} =
(\thrur{\partial} + {t}\thrul{V}_R\!({t} x) )P_R
+(\thrur{\partial} + {t} \thrul{V}_L\!({t} x) )P_L
+\mlr({t} x) P_R +\mrl({t} x) P_L
,
\end{equation}
and then expand the corresponding effective action in powers of the
parameter ${t}$. After extraction of a global factor $1/t^d$ in $d$
space-time dimensions, the terms of order $n$ contain $n$ covariant
derivatives (or gauge fields). At the diagrammatic level, a term of
order $n$ represents a one-loop Feynman graph with any number of
scalar fields and $n$ gauge fields as external legs, all of them at
zero four-momentum, or less gauge fields as external legs and
correspondingly more powers of the external momenta.  We emphasize
that gauge fields are assimilated to derivatives in such a way that
they are of the same order. This ensures preservation of gauge
invariance and as a consequence each order of the derivative expansion
of the effective action is separately invariant under gauge
transformations.

Several remarks can be made: i) In even-dimensional space-times and at
zero temperature, the derivative expansion of the effective action
contains even orders only. ii) The abnormal parity sector starts at
order $d$ in $d$ dimensions, since it contains the Levi-Civita
pseudo-tensor. iii) In $d$ dimensions, the terms beyond order $d$ are
ultraviolet (UV) convergent, and so free from UV ambiguities
introduced by the renormalization. iv) $\Gamma_{\text{gWZW}}$ contains
only terms of order $d$ (in $d$ dimensions). I.e., within the
derivative expansion, the WZW term vanishes at all orders beyond the
lowest order one. This means that $\Gamma^-=\Gamma_c^-$ except at
lowest order.  And v) within the derivative expansion, the
Minkowskian version of the effective action is real in the normal
parity and in the abnormal parity sectors. The derivative expansion is
an expansion around small external four-momentum, and so it can not
reach the analytical cuts related to particle production.

\subsection{The method of symbols}
\label{subsec:2.E}

A convenient technique to carry out calculations within the derivative
expansion is the method of symbols
\cite{Nepomechie:1984wt,Salcedo:1996qy}. The method has been extended
to curved space-time \cite{Vassilevich:2003xt,Salcedo:2006pv} and
finite temperature \cite{Garcia-Recio:2000gt,Megias:2002vr}.

This method will be used below in our calculation of the effective
action.

For any pseudo-differential operator $\hat f$ of the form $f(D,M)$,
constructed with covariant derivatives $D_\mu$ and external fields
$M(x)$ (all this non abelian in general) the method of symbols
states, for the diagonal matrix elements of $\hat f$,
\begin{equation}
\langle x |{\hat f}|x\rangle
=
\int\frac{d^d p }{(2\pi)^d} \langle x|f(D+p,M)|0\rangle
\label{eq:2.22}
\end{equation}
($d$ is the dimension of the space of $x$). $|0\rangle$ represents the
constant state $\langle x|0\rangle=1$, so that
$\partial_\mu|0\rangle=0$. For notational convenience, here and in
what follows, we use a purely imaginary momentum $p_\mu=i k_\mu$
($k_\mu$ real), but $p^2$ denotes $-p_\mu p_\mu=k^2$ and $d^dp=d^dk$.
The matrix element $\langle x|f(D+p,M)|0\rangle$ just coincides with
the standard symbol of $\hat f$, as defined in the theory of
pseudo-differential operators \cite{Seeley:1967ea}.

Invariance under the shift $p_\mu\to p_\mu + a_\mu$ in the momentum
integral (\ref{eq:2.22}) implies that $D_\mu$ can appear only in the
form $[D_\mu,~]$ in the final integrated expression. This ensures
gauge covariance of the right-hand side of (\ref{eq:2.22}),
consistently with the obvious gauge covariance of the left-hand side.

From (\ref{eq:2.22}) the derivative expansion is
easily obtained just by formally expanding in powers of
$D_\mu$.\footnote{$p_\mu$ is the momentum running in the quantum loop
  and is of order zero in the derivative expansion.  On the other hand
  $D_\mu$ now acts only on the other external fields present in $\hat
  f$ and counts as first order.  Direct expansion in powers of $D_\mu$
  in $f(D,M)$ (i.e. before applying the method of symbols) would not
  produce UV convergent integrals and would not correspond to the
  derivative expansion, as $D_\mu$ would contain not only the momenta
  of the external fields but also that of the quantum field running in
  the loop.}  The final step is to move all derivatives to the right
(derivating everything in passing as dictated by Leibniz's rule) and
to verify that terms with derivatives at the right (which would break
gauge invariance) vanish after carrying out the momentum integration.

The method of symbols then provides the (functional) trace of $\hat f$
as
\begin{equation}
\Tr (\hat f)
=
\int\frac{d^dx d^d p }{(2\pi)^d} \tr f(D+p,M)
.
\label{eq:2.23}
\end{equation}
(The brackets $\langle x|~|0 \rangle$ are usually omitted.) 

The method of symbols just described is simple enough but has the
drawback that gauge invariance is not manifest during the calculation.
An alternative technique, which we shall also employ in this work, is
that of {\em covariant} symbols \cite{Pletnev:1998yu,Salcedo:2006pv},
which has the virtue of being manifestly covariant from the beginning.
Quite simply, the method consists in applying a similarity
transformation in (\ref{eq:2.22}), which changes nothing:
\begin{equation}
\langle x |{\hat f}|x\rangle
=
\int\frac{d^d p }{(2\pi)^d} e^{-D_\mu \partial/\partial p_\mu}f(D+p,M)
e^{D_\nu \partial/\partial p_\nu}
\,.
\end{equation}
This can be seen to be equivalent to
\begin{equation}
\langle x |{\hat f}|x\rangle
=
\int\frac{d^d p }{(2\pi)^d} f(\bar{D},\bar{M})
\,,
\end{equation}
where $\bar{D}_\mu$ and $\bar{M}$ are manifestly gauge covariant:
\begin{eqnarray}
  \bar{M} &=& 
  M - [D_\alpha,M]\frac{\partial}{\partial p_\alpha} 
  + \frac{1}{2!}
[D_\alpha,[D_\beta,M]]\frac{\partial}{\partial p_\alpha} 
\frac{\partial}{\partial p_\beta} 
  -\cdots
\nonumber \\
  \bar{D}_\mu &=&  
  p_\mu - \frac{1}{2!}[D_\alpha,D_\mu]\frac{\partial}{\partial p_\alpha} 
+
\frac{2}{3!}[D_\alpha,[D_\beta,D_\mu]]
\frac{\partial}{\partial p_\alpha} \frac{\partial}{\partial p_\beta} 
  -\cdots
\label{eq:2.27}
\end{eqnarray}
and these series are truncated at the desired order in the derivative
expansion.

\section{Standard Model}
\label{sec:3}

\subsection{Fermion sector of the Standard  Model}
\label{subsec:3.A}

We can apply the previous general results to the Standard Model (SM) for
quarks and leptons coupled to gauge fields and Higgs 
\cite{Huang:1992bk,Amsler:2008zzb}. For quarks (and in Minkowski space)
\begin{equation}
{\cal L}_{\text{SM,q}}(x) =\bar{q}(x)\,i\Dirac_{\text{SM,q}} \, q(x) \,.
\end{equation}

In the notation of (\ref{eq:5}) the quark field, $q(x)$, is a column
matrix in the space 
${\cal H}_{LR}\otimes{\cal H}_{ud}\otimes
{\cal H}_{\text{gen}}\otimes{\cal H}_{\text{color}}\otimes
{\cal H}_{\text{Dirac}}$, where ${\cal H}_{LR} = 
{\cal H}_L\oplus {\cal H}_R$ (chirality) has dimension two,
${\cal H}_{ud}= {\cal H}_u\oplus {\cal H}_d$ 
($u$ or $d$ quark type) has dimension two,
${\cal H}_{\text{gen}}= {\cal H}_1\oplus {\cal H}_2\oplus {\cal H}_3$
(generation) has dimension three, ${\cal H}_{\text{color}}$ has
dimension three and ${\cal H}_{\text{Dirac}}$ has dimension four.
\begin{equation}
q(x)=
 \left(\begin{matrix}
 u_R  \\ d_R \\ u_L \\ d_L
\end{matrix}
\right),
\quad
\bar{q}(x)=
 \left(\begin{matrix}
 \bar{u}_L,   \bar{d}_L,   \bar{u}_R,  \bar{d}_R
\end{matrix}
\right) .
\end{equation}
Here only the $LR$ and $ud$ spaces are explicit while generation,
color and Dirac indices have been left implicit.
Likewise, in the unitary gauge, the Dirac operator takes the form
\begin{eqnarray}
i\Dirac_{\text{SM,q}}
&=&
\nonumber \\ &&
\hspace{-2cm}
\left(\begin{matrix}
-\frac{1}{\sqrt{2}} \phi Y_u 
&
0  
&
i\thrur{\partial} 
-\frac{1}{2}g\!\thrul{\tilde{W}}_3 
-\frac{1}{6}g^\prime\!\thru{\tilde{B}} 
-\frac{1}{2}g_s\lambda_a \!\thru{\tilde{G}}_a
&
-\frac{1}{\sqrt{2}}g \! \thrul{\tilde{W}}{}^+ 
\\
0 
&
-\frac{1}{\sqrt{2}} \phi Y_d 
&
-\frac{1}{\sqrt{2}}g\! \thrul{\tilde{W}}{}^-
&
i\thrur{\partial} 
+\frac{1}{2}g\!\thrul{\tilde{W}}_3 
-\frac{1}{6}g^\prime\!\thru{\tilde{B}} 
-\frac{1}{2}g_s\lambda_a \!\thru{\tilde{G}}_a
\\
 i\thrur{\partial} 
-\frac{2}{3}g^\prime\!\thru{\tilde{B}} 
-\frac{1}{2}g_s\lambda_a\! \thru{\tilde{G}}_a
&
0
&
-\frac{1}{\sqrt{2}} \phi Y_u^\dagger
&
0
\\
0
&
i\thrur{\partial} 
+\frac{1}{3}g^\prime\!\thru{\tilde{B}} 
-\frac{1}{2}g_s\lambda_a \!\thru{\tilde{G}}_a
&
0
&
-\frac{1}{\sqrt{2}} \phi Y_d^\dagger
\end{matrix}
\right)
.
\nonumber 
 \\
&&
\end{eqnarray} 

$Y_{u,d}$ are $3\times3$ matrices in generation space which denote the
Yukawa couplings of the quarks with the Higgs field $\phi(x)$.
$\tilde{G}_{a,\mu}$ are the gluon fields, with coupling constant
$g_s$, and $\lambda_a$ the Gell-Mann matrices in color
space. $\tilde{B}_\mu$ is the U(1) weak hypercharge gauge field, with
coupling constant $g^\prime$, $\tilde{W}^\pm_\mu$ and
$\tilde{W}_{3,\mu}$ are the SU(2) weak isospin gauge fields, with
coupling constant $g$. All matrices are the identity in generation
space, except $Y_{u,d}$, the identity in color space, except
$\lambda_a$, and the identity in Dirac space, except $\gamma^\mu$.

The gauge fields shown are the canonical ones. For convenience we shall
absorb the couplings in the fields, and write the Dirac operator in
the form
\begin{equation}
i\Dirac_{\text{SM,q}}
=
\left(\begin{matrix}
-\frac{\phi}{v} M_u 
&
0  
&
i\!\thru{D}_u 
-\thru{Z}
-\thru{G}
&
-\thrul{W}{}^+ 
\\
0 
&
-\frac{\phi}{v} M_d 
&
-\thrul{W}{}^-
&
i\!\thru{D}_d
+\thru{Z}
-\thru{G}
\\
i\!\thru{D}_u 
-\thru{G}
&
0
&
-\frac{\phi}{v} M_u^\dagger
&
0
\\
0
&
i\!\thru{D}_d
-\thru{G}
&
0
&
-\frac{\phi}{v} M_d^\dagger
\end{matrix}
\right)
.
\end{equation} 

In this expression $v$ is the vacuum expectation value of the Higgs
field after spontaneous symmetry breaking, and $M_{u,d}$ the complex
quark mass matrices,
\begin{equation}
v=\langle\phi\rangle , \quad
M_u =\frac{1}{\sqrt{2}} v  Y_u, \quad
M_d = \frac{1}{\sqrt{2}} v Y_d .
\end{equation}
\begin{equation}
G_\mu= 
\frac{1}{2}g_s\lambda_a \tilde{G}_{a,\mu} .
\end{equation}
Also
\begin{equation}
W^\pm_\mu= \frac{1}{\sqrt{2}}g \tilde{W}^\pm_\mu ,\quad
Z_\mu = \frac{1}{2}g\tilde{W}_{3,\mu}
-\frac{1}{2}g^\prime\tilde{B}_\mu
=
\frac{1}{2}\frac{g}{\cos\theta_W}\tilde{Z}_\mu ,
\end{equation}
where $\tilde{Z}_\mu(x)$ is the canonical field of the $Z^0$ boson and
$\theta_W$ the weak angle. In addition, we have introduced the
covariant derivatives
\begin{equation}
D_{u,\mu}= \partial_\mu + i A_{u,\mu} , \quad
D_{d,\mu}= \partial_\mu + i A_{d,\mu} , \quad
\end{equation}
with
\begin{equation}
A_{u,\mu}=
\frac{2}{3}g^\prime\tilde{B}_\mu , \quad
A_{d,\mu}=
-\frac{1}{3}g^\prime\tilde{B}_\mu .
\end{equation}

For leptons
\begin{equation}
{\cal L}_{\text{SM,l}}(x) =\bar{l}(x)\,i\Dirac_{\text{SM,l}} \, l(x) \,.
\end{equation}
The lepton field $l(x)$ belongs to the space
${\cal H}_{LR}\otimes{\cal H}_{\nu e}\otimes {\cal H}_{\text{gen}}
\otimes{\cal H}_{\text{Dirac}} $.  For convenience it includes a
spurious right-handed neutrino to achieve greater similarity with the
quark case. In matrix form the fields are organized as follows
\begin{equation}
l(x)=
 \left(\begin{matrix}
 \nu_R  \\ e_R \\ \nu_L \\ e_L
\end{matrix}
\right),
\quad
\bar{l}(x)=
 \left(\begin{matrix}
 \bar{\nu}_L,   \bar{e}_L,   \bar{\nu}_R,  \bar{e}_R
\end{matrix}
\right) ,
\end{equation}
and the Dirac operator takes the form (we assume massless neutrinos
throughout)
\begin{eqnarray}
i\Dirac_{\text{SM,l}}
&=&
\left(\begin{matrix}
0
&
0  
&
i\thrur{\partial} 
-\frac{1}{2}g\!\thrul{\tilde{W}}_3 
+\frac{1}{2}g^\prime\!\thru{\tilde{B}} 
&
-\frac{1}{\sqrt{2}}g \! \thrul{\tilde{W}}{}^+ 
\\
0 
&
-\frac{1}{\sqrt{2}}\phi Y_e
&
-\frac{1}{\sqrt{2}}g\! \thrul{\tilde{W}}{}^-
&
i\thrur{\partial} 
+\frac{1}{2}g\!\thrul{\tilde{W}}_3 
+\frac{1}{2}g^\prime\!\thru{\tilde{B}} 
\\
 i\thrur{\partial} 
&
0
&
0
&
0
\\
0
&
i\thrur{\partial} 
+g^\prime \!\thru{\tilde{B}} 
&
0
&
-\frac{1}{\sqrt{2}}\phi Y_e^\dagger
\end{matrix}
\right)
.
\end{eqnarray} 
The right-handed neutrino is completely decoupled. As for the quarks,
we find it convenient to rewrite the same matrix as
\begin{equation}
i\Dirac_{\text{SM,l}}
=
\left(\begin{matrix}
0
&
0  
&
i\!\thrur{\partial}
-\thru{Z}
&
-\thrul{W}{}^+ 
\\
0 
&
-\frac{\phi}{v} M_e
&
-\thrul{W}{}^-
&
i\!\thru{D}_e
+\thru{Z}
\\
i\!\thrur{\partial}
&
0
&
0
&
0
\\
0
&
i\!\thru{D}_e
&
0
&
-\frac{\phi}{v} M_e^\dagger
\end{matrix}
\right)
,
\end{equation} 
where we have introduced the new covariant derivative
\begin{equation}
D_{e,\mu}= \partial_\mu + iA_{e,\mu} , \quad
A_{e,\mu}=
-g^\prime\tilde{B}_\mu
\, .
\end{equation}

For subsequent use, we introduce the derivatives
\begin{eqnarray}
W^+_{\mu\nu} &=& 
D_{u,\mu} W^+_\nu - W^+_\nu D_{d,\mu} 
=
[\partial_\mu, W^+_\nu] + i (A_{u,\mu}- A_{d,\mu}) W^+_\nu
, \quad
\nonumber \\
W^-_{\mu\nu} &=& 
D_{d,\mu} W^-_\nu - W^-_\nu D_{u,\mu} 
=
[\partial_\mu, W^-_\nu] - i (A_{u,\mu}- A_{d,\mu}) W^-_\nu
,
\nonumber \\
F^{u,d}_{\mu\nu} &=& -i[D^{u,d}_\mu,D^{u,d}_\nu]
=
[\partial_\mu,A^{u,d}_\nu]-[\partial_\nu,A^{u,d}_\mu]
.
\label{3.10}
\end{eqnarray}
Note that the fields $W^\pm_{\mu\nu}$ {\em are not} antisymmetric in
$\mu$,$\nu$. Also note that the analogous construction in the lepton
sector gives exactly the same result as for the quark sector, namely,
$W^\pm_{\mu\nu} = [\partial_\mu, W^\pm_\nu] \pm i g^\prime\tilde{B}_\mu
W^\pm_\nu$. In fact, using the relation
\begin{equation}
A^{\text{e.m.}}_\mu = g^\prime\tilde{B}_\mu + 2\sin^2\theta_W\,Z_\mu
\quad\text{(photon field)}
\,,
\end{equation}
one finds
\begin{equation}
W^\pm_{\mu\nu} = D^{\text{e.m.}}_\mu W^\pm_\nu \mp
i2\sin^2\theta_W\,Z_\mu W^\pm_\nu \,,
\label{eq:3.17}
\end{equation}
which is covariant under $\text{U}_{\text{e.m.}}(1)$, the remaining
gauge freedom in the unitary gauge, apart from SU$_{\text{color}}$(3).

\subsection{Euclidean space}

We apply to the SM fields the prescriptions given in section
\ref{subsec:2.B} to go from Minkowskian to Euclidean space, and this
yields
\begin{eqnarray}
\Dirac_{\text{SM,q}}
&=&
\left(\begin{matrix}
\frac{\phi}{v} M_u 
&
0  
&
\thru{D}_u 
+\thru{Z}
+\thru{G}
&
\thrul{W}{}^+ 
\\
0 
&
\frac{\phi}{v} M_d 
&
\thrul{W}{}^-
&
\thru{D}_d
-\thru{Z}
+\thru{G}
\\
\thru{D}_u 
+\thru{G}
&
0
&
\frac{\phi}{v} M_u^\dagger
&
0
\\
0
&
\thru{D}_d
+\thru{G}
&
0
&
\frac{\phi}{v} M_d^\dagger
\label{3.18}
\end{matrix}
\right)
\\
\Dirac_{\text{SM,l}}
&=&
\left(\begin{matrix}
0
&
0  
&
\thrur{\partial}
+\thru{Z}
&
\thrul{W}{}^+ 
\\
0 
&
\frac{\phi}{v} M_e
&
\thrul{W}{}^-
&
\thru{D}_e
-\thru{Z}
\\
\thrur{\partial}
&
0
&
0
&
0
\\
0
&
\thru{D}_e
&
0
&
\frac{\phi}{v} M_e^\dagger
\end{matrix}
\right)
\end{eqnarray} 
with
\begin{equation}
D_{u,\mu}= \partial_\mu + A_{u,\mu} , \quad
D_{d,\mu}= \partial_\mu + A_{d,\mu} , \quad
D_{e,\mu}= \partial_\mu + A_{e,\mu}
\,.
\end{equation}

Also
\begin{eqnarray}
W^+_{\mu\nu} &=&  D_{u,\mu} W^+_\nu - W^+_\nu D_{d,\mu} 
, \quad
W^-_{\mu\nu} = 
D_{d,\mu} W^-_\nu - W^-_\nu D_{u,\mu} 
,
\nonumber \\
F^{u,d}_{\mu\nu} &=& [D^{u,d}_\mu,D^{u,d}_\nu]
.
\label{3.10a}
\end{eqnarray}

The c-number fields $Z_\mu$, $A_{u,\mu}$, $A_{d,\mu}$, $A_{e,\mu}$ and
$F^{u,d}_{\mu\nu}$ are purely imaginary, while $(W^+_\mu)^* =-W^-_\mu$
and $(W^+_{\mu\nu})^* =-W^-_{\mu\nu}$. The matrix field $G_\mu$ is
antihermitian. The field $\phi$ is a real c-number.

In what follows we shall work in Euclidean space, until section
\ref{sec:10}, where we return to Minkowskian space to display the
results.

\subsection{$ud$-parity}
\label{subsec:3.C}

In the SM only two of the four fields $Z_\mu$, $A_{u,\mu}$,
$A_{d,\mu}$, and $A_{e,\mu}$ are independent (and $F^d_{\mu\nu}$ is
proportional to $F^u_{\mu\nu}$), however, it proves useful to carry
out the calculation for the ``extended model'' where $Z_\mu$,
$A_{u,\mu}$, $A_{d,\mu}$, and $A_{e,\mu}$ are independent fields, and
also $M_u$ and $M_d$ are regarded as ($x$-independent) variables, as
this procedure provides simpler expressions. The quark sector with
generic $A_u$ and $A_d$ enjoys a $\text{U}_u(1)\times \text{U}_d(1)$
symmetry which in the SM reduces to $\text{U}_\text{e.m.}(1)$.
Moreover, for the extended model a symmetry becomes apparent under the
exchange of labels $u$ and $d$ in the quark sector, namely,
\begin{equation}
M_u\leftrightarrow M_d ,\quad
W^\pm\leftrightarrow W^\mp ,\quad
A_u\leftrightarrow A_d ,\quad
Z\leftrightarrow -Z ,\quad
G\leftrightarrow G ,\quad
\phi\leftrightarrow \phi .
\end{equation}
This corresponds to a similarity transformation of
$\Dirac_{\text{SM,q}}$ as given in (\ref{3.18}) and so it leaves the
(quark sector) effective action unchanged when expressed in terms of
generic $M_u$, $M_d$, $A_u$, $A_d$ and $Z$ (as well a $W^\pm$ and
$\phi$).  This symmetry, which we shall call $ud$-parity, is not
supported by the SM but it will be present in our calculation and this
will become useful later.

\section{CP violation}
\label{sec:4}

As noted, the full CP transformation in (\ref{eq:2.30}) is a symmetry
of the effective action functional. It is instructive to see this in
detail. First, note that, in Euclidean space and in four dimensions,
the definition of $\gamma_5$ does not contain an imaginary unit $i$,
and so no complex numbers are generated in the functional $\Gamma$
after taking Dirac traces (nor there are any other $i$'s in the Dirac
operator or the definition of $\Gamma$, cf.  (\ref{eq:2.28})). As a
consequence, when the background fields are replaced by their complex
conjugated, $\Gamma$ also becomes complex conjugated. That is, the
(real) normal parity component, $\Gamma^+$ is unchanged whereas the
(purely imaginary) abnormal parity component, $\Gamma^-$ changes to
$-\Gamma^-$. On the other hand, the transformation involving
$(x^0,\vec{x}) \to (x^0,-\vec{x})$ leaves invariant $\Gamma^+$ since
it does not contain $\epsilon_{\mu\nu\alpha\beta}$, but changes the
sign of $\Gamma^-$. In this way the complete effective action is left
invariant under full CP transformations.

Consider now the (Euclidean) {\em physical CP transformation} in the
SM. The transformation acts on the dynamical fields, namely, $G_a$,
$\phi$, $W^\pm$, $Z$, and $A_{u,d,e}$, but the constants $M_{u,d,e}$
are unchanged and in general this will no leave the effective action
invariant. This allows to classify the contributions to the effective
action in two types according to whether they are even or odd under
physical CP:
\begin{equation}
\Gamma=\Gamma_+ + \Gamma_-
\label{eq:4.1}
\,.
\end{equation}
(This classification is not to be confused with the separation
(\ref{eq:2.8}) into terms even and odd under pseudo-parity.)

In view of the fact that the full CP transformation leaves $\Gamma$
invariant, it follows that the physical CP transformation has the same
effect on the effective action as the transformation ($v$ is real):
\begin{equation}
M_{u,d,e}\to M_{u,d,e}^*
\,.
\label{eq:2.35}
\end{equation}
This implies the well known result that if the Yukawa couplings were
real, or equivalent to real, there would be no CP violation.  This is
automatically the case in the lepton sector but not in the quark
sector for three or more generations \cite{Kobayashi:1973fv}. Indeed,
for arbitrary complex matrices $M_u$ and $M_d$ one can write
\begin{equation}
M_u={\cal A}_{u,L}^{-1}m_u {\cal A}_{u,R}\,,\quad
M_d={\cal A}_{d,L}^{-1}m_d {\cal A}_{d,R}\,,
\end{equation}
where the matrices $m_u$, $m_d$ are diagonal and non negative and
${\cal A}_{u,L}$, ${\cal A}_{u,R}$, ${\cal A}_{d,L}$, ${\cal A}_{d,R}$
are unitary, all of them in generation space. Using the freedom to
rotate the quark fields in generation space allows to bring the Dirac
operator to the form
\begin{eqnarray}
\Dirac_{\text{SM,q}}
&=&
\left(\begin{matrix}
\frac{\phi}{v} \Omega_1 C^{-1} m_u \,\Omega_2 
&
0  
&
\thru{D}_u 
+\thru{Z}
+\thru{G}
&
\thrul{W}{}^+ 
\\
0 
&
\frac{\phi}{v} \Omega_1  m_d \,\Omega_3
&
\thrul{W}{}^-
&
\thru{D}_d
-\thru{Z}
+\thru{G}
\\
\thru{D}_u 
+\thru{G}
&
0
&
\frac{\phi}{v} \Omega_2^{-1} m_u C \Omega_1^{-1}
&
0
\\
0
&
\thru{D}_d
+\thru{G}
&
0
&
\frac{\phi}{v} \Omega_3^{-1} m_d \,\Omega_1^{-1}
\end{matrix}
\right)
\end{eqnarray} 
where
\begin{equation}
C= {\cal A}_{u,L}{\cal A}_{d,L}^{-1}
\end{equation}
is the Cabibbo-Kobayashi-Maskawa (CKM) matrix, and $\Omega_{1,2,3}$
are arbitrary unitary matrices (in generation space).  It is manifest,
by means of an appropriate choice of $\Omega_{1,2,3}$, that the
physics is invariant under the redefinition $C\to U_1 C U_2$ where
$U_{1,2}$ are arbitrary unitary and diagonal matrices (so that they
commute with $m_u$ and $m_d$).  On the other hand, choosing
$\Omega_{1,2,3}$ as the identity matrix simplifies the Dirac operator
and shows that $C$ has to be complex to allow violation of physical
CP; the CP transformation as given in (\ref{eq:2.35}) becomes
equivalent to
\begin{equation}
C\to C^*
.
\end{equation}

The similar manipulations in the lepton sector do not give rise to a
 matrix $C$ since the would-be $M_\nu$ complex mass matrix vanishes. So
the lepton sector does not contribute to CP violation in the effective
action and this sector will be disregarded in what follows.

Alternatively, the quark rotations in generation space can be chosen
so that the Dirac operator becomes
\begin{eqnarray}
\Dirac_{\text{SM,q}}
&=&
\left(\begin{matrix}
\frac{\phi}{v} m_u 
&
0  
&
\thru{D}_u 
+\thru{Z}
+\thru{G}
&
\thrul{W}{}^+ C
\\
0 
&
\frac{\phi}{v} m_d
&
\thrul{W}{}^- C^{-1}
&
\thru{D}_d
-\thru{Z}
+\thru{G}
\\
\thru{D}_u 
+\thru{G}
&
0
&
\frac{\phi}{v} m_u
&
0
\\
0
&
\thru{D}_d
+\thru{G}
&
0
&
\frac{\phi}{v} m_d
\end{matrix}
\right)
.
\end{eqnarray} 
This shows that only contributions involving $W^\pm$ can appear in
the CP violating sector, $\Gamma_-$.

Because all quantities in the Dirac operator are c-numbers in
generation space except the Yukawa couplings, and the effective action
adds one-loop graphs with quarks running on them, this functional can
be arranged in the form
\begin{equation}
\Gamma= \sum_{\lambda}\tr f_\lambda(M_u,M_d) \int d^4x\,
\tr {\cal O}_\lambda(x)
\,,
\label{eq:4.8}
\end{equation}
where $f_\lambda(M_u,M_d)$ are operators in generation space
constructed with the complex mass matrices and ${\cal O}_\lambda$ are
local operators constructed with the Higgs and the various gauge
fields. In the first case the trace refers to generation space, ${\cal
  H}_{\text{gen}}$, and it refers to all the other spaces in the
second case.  As noted before all these operators do not involve any
complex number in their construction in terms of the fields and mass
matrices.  As a consequence, the CP transformation (\ref{eq:2.35})
implies
\begin{equation}
\tr f_\lambda(M_u,M_d)
\to
\tr f_\lambda(M_u^*,M_d^*)
=(\tr f_\lambda(M_u,M_d))^*
,
\end{equation}
therefore the CP violating component of the effective action can be
expressed as
\begin{equation}
\Gamma_-= \sum_{\lambda}i\,\Im (\tr f_\lambda(M_u,M_d)) \int d^4x\,
\tr {\cal O}_\lambda(x)
\,.
\label{eq:4.10}
\end{equation}
It also follows that the local operators ${\cal O}(x)$ contributing to
$\Gamma^+_-$ (normal parity, CP violating) are antihermitian, since
$\Gamma^+$ is real in Euclidean space. On the other hand the local
operators in $\Gamma^-_-$ (abnormal parity, CP violating) must be
hermitian.  In this reasoning we use the fact that the two
transformations involved, namely, CP (which defines the separation in
(\ref{eq:4.1})) and pseudo-parity (which defines the separation in
(\ref{eq:2.8})) commute. This is correct since pseudo-parity just
exchanges the labels $L$ and $R$ while CP does not mix those labels.

An important remark is that in the Standard Model all bosons can be
assigned natural parity, $(-1)^J$. Therefore the abnormal parity
sector is just the P odd sector, while the normal parity sector of the
CP odd component is C odd and P even. This is unlike the chirally
broken phase of QCD where parity is preserved even by the abnormal
parity sector due to the presence of abnormal parity hadrons.

\section{Momentum integrals}
\label{sec:5}

As we shall see, for the operators $\tr f_\lambda(M_u,M_d)$ depending
on the complex mass matrices, cf. (\ref{eq:4.8}), the derivative
expansion produces the set of integrals
\begin{eqnarray}
I^k_{r_1,t_1,\ldots,r_n,t_n}
&=&
\nonumber \\
&&
\hskip -2cm
\int \frac{d^4p}{(2\pi)^4} (p^2)^k
\,\tr
\!\!
\left[
\frac{1}{(p^2+M_u M_u^\dagger)^{r_1}}
\frac{1}{(p^2+M_d M_d^\dagger)^{t_1}}
\cdots
\frac{1}{(p^2+M_u M_u^\dagger)^{r_n}}
\frac{1}{(p^2+M_d M_d^\dagger)^{t_n}}
\right]
\nonumber \\
&&
\hskip -2cm
= \int \frac{d^4p}{(2\pi)^4} (p^2)^k
\,\tr
\!\!
\left[
\frac{1}{(p^2+m_u^2)^{r_1}}C
\frac{1}{(p^2+m_d^2)^{t_1}}
C^{-1}
\cdots
\frac{1}{(p^2+m_u^2)^{r_n}}C
\frac{1}{(p^2+m_d^2)^{t_n}}
C^{-1}
\right]
\nonumber \\
&&
\label{eq:6.1}
\end{eqnarray}
where the exponents $k$ and $r_i$, $t_i$ are non negative integers.
The integral with $2n$ indices will appear in contributions with $n$
$W^+$ and $n$ $W^-$. (Charge conservation requires as many $W^+$ as
$W^-$ in any contribution to the effective action and $W^\pm$ are the
only fields connecting the spaces $u$ and $d$.)

Due to the cyclic property of the trace
\begin{equation}
I^k_{r_1,t_1,\ldots,r_n,t_n} = I^k_{r_2,t_2,\ldots,r_n,t_n,r_1,t_1}
,
\end{equation}
and also taking, the hermitian adjoint,
\begin{equation}
(I^k_{r_1,t_1,\ldots,r_n,t_n})^*
 = I^k_{r_n,t_{n-1},r_{n-1},\ldots,t_1,r_1,t_n}
.
\label{eq:6.3}
\end{equation}

Since we have seen above that the CP violating component is tied to
the imaginary part of this integral, we introduce the definition
\begin{equation}
\hat{I}^k_{r_1,t_1,\ldots,r_n,t_n}
= i\,\Im I^k_{r_1,t_1,\ldots,r_n,t_n}
,
\end{equation}
which enjoys the properties
\begin{equation}
\hat{I}^k_{r_1,t_1,\ldots,r_n,t_n} = \hat{I}^k_{r_2,t_2,\ldots,r_n,t_n,r_1,t_1}
 = 
-\hat{I}^k_{r_n,t_{n-1},r_{n-1},\ldots,t_1,r_1,t_n}
\,.
\end{equation}

From these relations it is immediate that $\hat{I}$ vanishes if $n=0$
or $n=1$. Therefore, at least 2 $C$ and 2 $C^{-1}$ are needed to have
a contribution to $\Gamma_-$, or equivalently, 2 $W^+$ and 2 $W^-$.
This is a well known fact in the literature \cite{Jarlskog:1985ht}.
(This implies that the operator $\tr(F_{\mu\nu}\tilde{F}^{\mu\nu})$
\cite{Shaposhnikov:1987pf} mentioned in the Introduction cannot be
derived from simple integration of the quarks as the would-be term
with four $W^\pm$ vanishes. Such term can be produced if internal
gauge field lines are allowed.)

Of particular interest will be the first non trivial case, $n=2$. For
it one finds
\begin{equation}
\hat{I}^k_{r_1,t_1,r_2,t_2}
 = 
-\hat{I}^k_{r_2,t_1,r_1,t_2}
 = 
-\hat{I}^k_{r_1,t_2,r_2,t_1}
\,.
\label{eq:5.6}
\end{equation}

It will also be useful to note the transformation of the momentum
integral under $ud$-parity (see section \ref{subsec:3.C}), namely,
\begin{equation}
I^k_{r_1,t_1,\ldots,r_n,t_n}
\longrightarrow 
I^k_{t_1,\ldots,r_n,t_n,r_1}
\,,
\end{equation}
and in particular
\begin{equation}
\hat{I}^k_{r_1,t_1,r_2,t_2}
\longrightarrow 
-\hat{I}^k_{t_1,r_1,t_2,r_2}
\, .
\end{equation}

Using the second form in (\ref{eq:6.1}), to compute the integrals of
the type $\hat{I}^k_{r_1,t_1,r_2,t_2}$ for three generations, one can
use the identity \cite{Jarlskog:1985ht}
\begin{equation}
\Im(C_{ij}C^{-1}_{jk}C_{kl}C^{-1}_{li})
=
J\epsilon_{ik}\epsilon_{jl}
,\quad
i,j,k,l=1,2,3
\,,
\label{eq:5.9}
\end{equation}
(with no implicit summation over repeated indices here) where
$\epsilon_{ij}=\sum_{k=1}^3\epsilon_{ijk}$, and $J$ is the Jarlskog
invariant \cite{Jarlskog:1985ht,Amsler:2008zzb}
\begin{equation}
  J=\cos\theta_{12}\,\cos^2\theta_{13}\,\cos\theta_{23}\,
\sin\theta_{12}\,\sin\theta_{13}\,\sin\theta_{23}\,\sin\delta 
= 3.0(2)\times 10^{-5}
  .
\end{equation}
The resulting momentum integrals no longer involve matrices in the
integrand.

The only integral required in this paper is $\hat{I}_{1,1,2,2}^3$. It
can be cast in the form
\begin{equation}
\hat{I}_{1,1,2,2}^3
=
iJ \,G_F \,\kCP
\,.
\label{eq:5.11}
\end{equation}
$\kCP$ is a dimensionless coefficient and $G_F$ is the Fermi constant,
which can be related to the the vacuum expectation value of the Higgs
field, $G_F=1/(\sqrt{2} v^2)$.  For $v$ and for the quark masses we
take $v=246\,\text{GeV}$, $m_u=2.55\,{\rm MeV}$, $m_d=5.04\,{\rm
  MeV}$, $m_s=104\,{\rm MeV}$, $m_c=1.27\,{\rm GeV}$, $m_b=4.2\,{\rm
  GeV}$, $m_t=171.2\,{\rm GeV}$ \cite{Amsler:2008zzb}.  This gives
\begin{equation}
\kCP= 3.1\times 10^2
\,,
\qquad
G_F\,\kCP= 3.6\times 10^{-3}\,\text{GeV}^{-2}
\,.
\end{equation}
 
The integral $\hat{I}_{1,1,2,2}^3$ is an homogeneous function of the
quark masses of degree $-2$ so $\kCP$ can be expressed as a function
of the Yukawa couplings, $y_q=\sqrt{2}m_q/v$.  The loop function
$\kCP$ is of great interest by itself so we give full details of its
form and calculation in section \ref{sec:12}. We only remark here that
it is numerically many orders of magnitude larger than the ``minimal''
term $(y_u^2-y_c^2)(y_c^2-y_t^2)(y_t^2-y_u^2)(y_d^2-y_s^2)%
(y_s^2-y_b^2)(y_b^2-y_d^2)=6\times 10^{-18}$. The reason, of course,
is that other non minimal factors appear in the full expression.

\section{Derivative expansion in the SM}
\label{sec:6}

Smit \cite{Smit:2004kh} has proposed to use the derivative expansion
as a suitable approach in the present context of CP violation in the
SM.

In the SM each $W^\pm$, $Z$, $D_{u,d,\nu}$ or $G_a$ counts as first
order. As we have seen, at least four $W^\pm$ are needed in the CP
violating sector, therefore $\Gamma_-$ vanishes at zeroth or second
order in the derivative expansion. It was shown in \cite{Smit:2004kh}
that the fourth order is also vanishing in the abnormal parity sector.
We verify below that the fourth order vanishes actually in both
sectors. Therefore, as suggested in \cite{Smit:2004kh}, the first non
trivial contribution should start at six derivatives. Hernandez et al.
\cite{Hernandez:2008db} have addressed such a computation in the
abnormal parity sector and find a non vanishing result. We compute
below all contributions to sixth order, including Higgs and normal
parity terms. Our result do not sustain those in
\cite{Hernandez:2008db}. We find non vanishing contributions in the normal
parity sector but none in the abnormal parity one.

As we have noted above we do not need to consider leptons since they
do not give a contribution to $\Gamma_-$ assuming massless neutrinos.
On the other hand we can also neglect gluons. At fourth order the four
$W^\pm$ already saturate the required number of derivatives and no
gluons are allowed. At sixth order there is room for up to two gluon
fields. However, by gauge invariance one gluon is not admissible and
two gluons must be combined to form a gluon field strength. Such term
vanishes due to the trace on color. In what follows gluons are not
included and color just gives a global factor $N_c=3$.

Another question is whether the derivative expansion at low orders
produces a reliable approximation to the physical amplitudes. Using
simple estimates, it has been argued in \cite{Hernandez:2008db} that
the range of validity could reach the scale of the charm quark mass or
even larger. Besides, it is clearly of interest to have correctly
accounted for the lowest order operators of the effective action, as
guidance on the available CP violating mechanisms.

\section{Chiral invariant approach to the effective action}
\label{sec:7}

The approach of section \ref{sec:2.D} can be directly applied to the SM.

For the SM in the quark sector, the operator $K$ of (\ref{eq:2.20})
takes the form (the gluons are no longer present in the covariant
derivatives)
\begin{equation}
K =
\left(
 \begin{matrix}
(\phi^2/ v^2)
M_u M_u^\dagger - (\thru{D}_u + \thru{Z})
(\thru{D}_u + \thru{\varphi})  &
  -\thrul{W}{}^+
(\thru{D}_d + \thru{\varphi})
\cr
 -\thrul{W}{}^-
(\thru{D}_u + \thru{\varphi})  &
(\phi^2/ v^2)
M_d M_d^\dagger - 
(\thru{D}_d - \thru{Z})
(\thru{D}_d +  \thru{\varphi})
\end{matrix}
\right)
.
\label{eq:7.1}
\end{equation}
Here we have introduced the shorthand notation
\begin{equation}
\varphi_\mu(x)= \phi^{-1}[\partial_\mu,\phi] .
\label{eq:7.2}
\end{equation}
The operator $K$ acts in the space ${\cal H}_{ud}\otimes {\cal
  H}_{\text{gen}}\otimes{\cal H}_{\text{color}}\otimes {\cal
  H}_{\text{Dirac}}$ (that is, the same space as
$\Dirac_{\text{SM,q}}$ except the factor ${\cal H}_{LR}$).  Therefore,
for the SM the equations (\ref{eq:2.21}) become
\begin{eqnarray}
\Gamma^+
&=&
-\frac{1}{2}N_c \,\Re\langle \log K\rangle
,
\nonumber \\
\Gamma^-
&=&
-\frac{1}{2}N_c \, i\,\Im \langle\log K\rangle
+\Gamma_{\text{gWZW}}
.
\label{eq:8.6}
\end{eqnarray}

Here we have introduced the symbol $\langle~\rangle$ which will be
used in what follows. It denotes a trace operation {\em including a
  $\gamma_5$ in the abnormal parity sector}, and just the trace,
without $\gamma_5$, in the normal parity sector. The precise trace
operation implied by $\langle~\rangle$ will often be obvious from the
context. The inclusion of $\gamma_5$ does not spoil the cyclic
property for $\langle~\rangle$ since all operators involved will have
an even number of Dirac gamma matrices. In this way we can treat
simultaneously the normal and abnormal parity components.

In (\ref{eq:8.6}) the trace implied by $\langle~\rangle$ acts on
$x$-space and on ${\cal H}_{ud}\otimes {\cal H}_{\text{gen}}\otimes
{\cal H}_{\text{Dirac}}$. 

It has been shown in \cite{Smit:2004kh} that the term
$\Gamma_{\text{gWZW}}$ does not have a contribution to the CP
violating component of the effective action, so we disregard this term
in what follows.

For convenience, let us separate $K$ into its diagonal and off
diagonal parts (in $ud$ space)
\begin{eqnarray}
K &=& K_D+K_A
,
\nonumber \\
K_D &=&
\left(
 \begin{matrix}
(\phi^2/ v^2)
M_u M_u^\dagger - (\thru{D}_u + \thru{Z})
(\thru{D}_u  + \thru{\varphi} )  &
0
\cr
0 &
(\phi^2/ v^2)
M_d M_d^\dagger - 
(\thru{D}_d - \thru{Z})
(\thru{D}_d + \thru{\varphi})
\end{matrix}
\right)
,
\nonumber \\
K_A &=&
\left(
 \begin{matrix}
0 &
  -\thrul{W}{}^+
(\thru{D}_d + \thru{\varphi} )
\cr
 -\thrul{W}{}^-
(\thru{D}_u + \thru{\varphi})  &
0
\end{matrix}
\right)
.
\label{eq:7.4}
\end{eqnarray}

The advantage of this separation is that only $K_A$ contains charged
currents, which can break CP, and also that this term is of first
order in derivatives\footnote{$K_A$ is of order one and not of order
  two because the counting refers to derivatives of the external
  fields and in (\ref{eq:7.4}) $\thru{D}_{u,d}$ can still act on the
  quarks. As noted previously in section \ref{subsec:2.E}, such
  derivatives on the running fermion are of order zero in the
  derivative counting.} (recall that the gauge fields, and in
particular $W^\pm$, count as first order) so the $n$-th order in the
derivative expansion of $\Gamma$ can contain at most $n$ factors $K_A$.

Substituting this form in $\langle \log K\rangle$, which appears in
(\ref{eq:8.6}), yields
\begin{eqnarray}
\llangle \log K\rrangle
&=&
\llangle \log K_D\rrangle
-\frac{1}{2}
\llangle \frac{1}{K_D}K_A\frac{1}{K_D}K_A\rrangle
-\frac{1}{4}
\llangle \frac{1}{K_D}K_A\frac{1}{K_D}K_A\frac{1}{K_D}K_A
\frac{1}{K_D}K_A\rrangle
-
\cdots
\nonumber \\
&=&
\sum_{n=0}^\infty 
\llangle \log K\rrangle_{2n}
.
\end{eqnarray}
The subindex $2n$ in $\llangle \log K\rrangle_{2n}$ indicates that the
term contains exactly $2n$ $W^\pm$ fields. Working out the trace in
$ud$ space (for $n>0$) one obtains
\begin{eqnarray}
\llangle \log K\rrangle_{2n}
&=&
-\frac{1}{n}\Bigg\langle
\Bigg[
\left(
\frac{\phi^2}{ v^2}
M_u M_u^\dagger - (\thru{D}_u + \thru{Z})
(\thru{D}_u + \thru{\varphi})
\right)^{-1}
  \thrul{W}{}^+
(\thru{D}_d + \thru{\varphi})
\nonumber \\
&&\times
\left(
\frac{\phi^2}{ v^2}
M_d M_d^\dagger - 
(\thru{D}_d - \thru{Z})
(\thru{D}_d + \thru{\varphi} )
\right)^{-1}
\thrul{W}{}^-
(\thru{D}_u + \thru{\varphi})
\Bigg]^n
\Bigg\rangle
.
\label{eq:7.5}
\end{eqnarray}
Here the trace implied by $\llangle~\rrangle$ refers to ${\cal
  H}_{\text{gen}}\otimes {\cal H}_{\text{Dirac}}$, and $x$-space. This
expression is manifestly $ud$-parity invariant (see section
\ref{subsec:3.C}), as can be verified by using the trace cyclic property.

Because at least four $W^\pm$ are required to have a CP violating
term, the relevant contributions start at $\llangle \log K\rrangle_4$.
To sixth order in the derivative expansion only $\llangle \log
K\rrangle_4$ and $\llangle \log K\rrangle_6$ have to be retained.
$\llangle \log K\rrangle_4$ contains terms with at least four
derivatives (namely, those coming from $W^\pm$), while $\llangle \log
K\rrangle_6$ starts at six derivatives:
\begin{eqnarray}
\llangle \log K\rrangle_4 
&=& 
\llangle \log K\rrangle_{4+0} + \llangle \log K\rrangle_{4+2} + {\cal O}(D^8) ,
\nonumber \\
\llangle \log K\rrangle_6 
&=& 
\llangle \log K\rrangle_{6+0} + {\cal O}(D^8) 
.
\end{eqnarray}
Here $\llangle \log K\rrangle_{2n+2m}$ indicates $2n$ derivatives from
$W^\pm$ and $2m$ more derivatives not from $W^\pm$, i.e., coming from
$\partial_\mu$, $Z_\mu$ or $A^{u,d}_\mu$.

\section{Vanishing of terms of the type  $2n+0$}
\label{sec:8}

In Ref.~\cite{Smit:2004kh}, and also confirmed in
\cite{Hernandez:2008db}, it was shown that there is no CP violating
contribution to four derivatives in the abnormal parity sector. Let us
show that this is true in both sectors and that, moreover, there is no
sixth order contribution either coming from terms with six $W^\pm$.

These statements are remarkably easy to establish using the method of
symbols described in section \ref{subsec:2.E}.  This method amounts to
make the replacement $\thru{D}_{u,d} \to \thru{D}_{u,d} + \thrur{p}$
in $K_D$ and $K_A$ and integrate over $p_\mu$. (Recall that $p_\mu$ is
purely imaginary but $p^2=-p_\mu^2$, as explained in section
\ref{subsec:2.E}.)

Concretely, the CP violating terms with precisely four derivatives
must come from $\llangle \log K\rrangle_4$ taking no other derivatives
than the four $W^\pm$ (i.e., must be of the type $4+0$).  Therefore,
to four derivatives, we can set $\thru{D}_{u,d} \to \thrur{p}$,
$\thru{Z}\to 0$ and $\thru{\varphi} \to 0$ in the operator $K$:
\begin{eqnarray}
\llangle \log K\rrangle_{4+0}
&=&
-\frac{1}{2}
\int\frac{d^4 x d^4 p }{(2\pi)^4}
\llangle
\Bigg[
\left(
\frac{\phi^2}{ v^2}
M_u M_u^\dagger +p^2 \right)^{-1}
  \thrul{W}{}^+ \thrur{p}
\left(
\frac{\phi^2}{ v^2}
M_d M_d^\dagger + p^2
\right)^{-1}
\thrul{W}{}^-
\thrur{p}
\Bigg]^2
\rrangle
.
\nonumber \\ &&
\label{eq:9.3}
\end{eqnarray}

Here it is already obvious that, upon momentum integration, the
integral $I^2_{1,1,1,1}$ (introduced in (\ref{eq:6.1})) will be
generated. Because this integral is real, due to eq.~(\ref{eq:6.3}),
it follows (cf. (\ref{eq:4.10})) that no CP violating term is produced
to fourth order in the derivative expansion, neither in the normal nor
the abnormal parity sectors.

To see this in more detail, we first take an angular average in
(\ref{eq:9.3}), using 
\begin{equation}
p_\mu p_\nu p_\alpha p_\beta \to
(\delta_{\mu\nu}\delta_{\alpha\beta}+\delta_{\mu\alpha}\delta_{\nu\beta}+
\delta_{\mu\beta}\delta_{\alpha\nu})p^4/(d(d+2))
.
\end{equation}
Since no derivatives with respect to $x$ are present in the expression,
we can simply rescale $p_\mu \to (\phi(x)/v)p_\mu$, and the momentum
integrals in (\ref{eq:6.1}) apply. Specifically,
\begin{eqnarray}
\llangle \log K\rrangle_{4+0}
&=&
-\frac{1}{48}
I^2_{1,1,1,1}
\int d^4 x
\,
\big\langle
\left(
  \thrul{W}{}^+ \gamma_\mu
\thrul{W}{}^- \gamma_\mu
\right)^2
+
\left(
  \thrul{W}{}^+ \gamma_\mu
\thrul{W}{}^- \gamma_\nu
\right)^2
\nonumber \\ && 
+
  \thrul{W}{}^+ \gamma_\mu
\thrul{W}{}^- \gamma_\nu
  \thrul{W}{}^+ \gamma_\nu
\thrul{W}{}^- \gamma_\mu
\big\rangle
.
\end{eqnarray}
The result to four derivatives is proportional to $I^2_{1,1,1,1}$, as
advertised.

If we consider now the case of sixth order $6+0$, i.e., when the six
derivatives are saturated by six $W^\pm$, it is quite clear, by using
the same reasoning, that the result will be proportional to
$I^2_{1,1,1,1,1,1}$, which is also real, and therefore, also no CP
violation is produced in either sector from such contributions.  Of
course, the analogous result holds for all orders of the type $2n+0$
too. That is,
\begin{equation}
(\Gamma_-)_{2n+0}=0
\,.
\end{equation}

We note that this vanishing is rather trivial in the abnormal
parity sector (for space-time dimension $d>2$): with the Levi-Civita
tensor and only two four-vectors $W^\pm_\mu$ it is not possible to
construct a non vanishing scalar.

The vanishing for the normal parity part is also easily understood.
Due to charge conservation, the possible operators constructed using
only $W^\pm_\mu$ are of the type \\
$((W^+_\mu W^+_\mu) (W^-_\nu
W^-_\nu))^n (W^+_\alpha W^-_\alpha)^m$ and are CP even.

\section{CP violation in the absence of Higgs field derivatives}
\label{sec:9}

We have just seen that no CP violation occurs to four derivatives and
also to six derivatives if these are saturated by $W^\pm$'s. Therefore,
any CP violation through sixth order in the derivative expansion must
come from terms with four $W^\pm$ and two other derivatives not of the
$W^\pm$ type, that is, terms of the type $4+2$:
\begin{equation}
\Gamma_- =  (\Gamma_-)_{4+2} + {\cal O}(D^8)
\,.
\end{equation}

In the next section we shall consider the general case. Presently, we
study the simplest situation where no derivatives of the Higgs field
are considered. In this case, the Higgs field $\phi(x)$ itself can be
set to its vacuum expectation value $v$, since it can be restored in
the formulas at the end by a rescaling of the quark masses. Under
these assumptions, the trace $\llangle \log K\rrangle_4$ of
(\ref{eq:7.5}) reduces to the simpler form
\begin{equation}
\llangle \log K\rrangle_4
=
-\frac{1}{2}\Big\langle
\Big[
\left(
M_u M_u^\dagger 
- (\thru{D}_u + \thru{Z})
\thru{D}_u \right)^{-1}
  \thrul{W}{}^+
\! \thru{D}_d
\left(
M_d M_d^\dagger - 
(\thru{D}_d - \thru{Z})
\thru{D}_d \right)^{-1}
\thrul{W}{}^-
\! \thru{D}_u
\Big]^2
\Big\rangle
.
\end{equation}

Here we apply once again the method of symbols making the replacement
$D_\mu\to p_\mu+D_\mu$, integrating over $p_\mu$ and then expanding in
powers of the derivatives:
\begin{equation}
\llangle \log K\rrangle_4
=
-\frac{1}{2}
\int\frac{d^4 x d^4 p }{(2\pi)^4}
\Big\langle
\tilde{N}_u 
  \thrul{W}{}^+
 (\thrur{p} + \thru{D}_d )
\tilde{N}_d 
\thrul{W}{}^-
(\thrur{p} +  \thru{D}_u)
\tilde{N}_u 
  \thrul{W}{}^+
 (\thrur{p} + \thru{D}_d )
\tilde{N}_d 
\thrul{W}{}^-
(\thrur{p} +  \thru{D}_u)
\Big\rangle
.
\label{9.2}
\end{equation}
Where
\begin{equation}
\tilde{N}_u 
= 
\big(
M_u M_u^\dagger 
- (\thrur{p} + \thru{D}_u + \thru{Z})
(\thrur{p} + \thru{D}_u) 
\big)^{-1}
,
\quad
\tilde{N}_d 
= 
\big(
M_d M_d^\dagger 
- (\thrur{p} + \thru{D}_d - \thru{Z})
(\thrur{p} + \thru{D}_d) 
\big)^{-1}
.
\end{equation}

Expansion of the first denominator gives
\begin{equation}
\tilde{N}_u 
= 
N_u 
+  
N_u^2 \big( 
\thrur{p} \, \thru{D}_u + 
(\thru{D}_u + \thru{Z} )\thrur{p}
+ 
(\thru{D}_u + \thru{Z} ) \thru{D}_u 
\big) 
+  
N_u^3 \big( 
\thrur{p} \, \thru{D}_u + 
(\thru{D}_u + \thru{Z} )\thrur{p}
\big)^2 
+  {\cal O}(D^3)
\,.
\label{9.4}
\end{equation}
and similarly for $\tilde{N}_d$. And we have defined
\begin{equation}
N_u
=
\big(
 M_u M_u^\dagger + p^2
\big)^{-1}
,
\quad
N_d
=
\big(
 M_d M_d^\dagger + p^2
\big)^{-1}
\,.
\end{equation}
The two objects $N_u$ and $N_d$ are $x$-independent, they do not
commute with each other but commute with all other quantities in
$\llangle \log K\rrangle_4$. In addition, they appear in the momentum
integrals introduced in section \ref{sec:5}.  Indeed, the first eq. in
(\ref{eq:6.1}) can be rewritten as
\begin{equation}
I^k_{r_1,t_1,\ldots,r_n,t_n}
=
\int \frac{d^4p}{(2\pi)^4} (p^2)^k
\,\tr
\!\!
\left[
N_u^{r_1} N_d^{t_1}
\cdots
N_u^{r_n} N_d^{t_n}
\right]
.
\end{equation}

At this point considerable simplification can be achieved by making
the following observation. To produce $\llangle \log K\rrangle_{4+2}$
from (\ref{9.2}) we need to pick up exactly two derivatives (apart
from the four explicit $W^\pm$). On the other hand, CP violating
contributions come only from the imaginary part of
$I^k_{r_1,t_1,r_2,t_2}$, and this requires $r_1\not=r_2$ and
$t_1\not=t_2$, cf. (\ref{eq:5.6}).  Now, it is quite clear that, in
order to obtain such a situation, it is necessary to pick up exactly
one of the derivatives from one of the $\tilde{N}_u$ and the other
derivative from one of the $\tilde{N}_d$.  Any other possibility ends
up with either the two $N_u$ or the two $N_d$ raised to the same
power, i.e., $r_1=r_2=1$ or $t_1=t_2=1$. Therefore the CP violating
contributions in the present case come from $I^k_{1,1,2,2}$.

To work this out let us simplify the expressions by introducing the
following quantities, which appear naturally in (\ref{9.2}) and
(\ref{9.4})
\begin{eqnarray}
\w^\pm=\thrul{W}{}^\pm \! \thrur{p}
,
\quad
\delta_u = 2 p D_u + \thru{Z}\,\thrur{p}
,
\quad
\delta_d = 2 p D_d - \thru{Z}\,\thrur{p}
.
\end{eqnarray}
Then, applying the previous observation yields
\begin{eqnarray}
\llangle \log K\rrangle_{4+2}
&=&
-\frac{1}{2}
\int\frac{d^4 x d^4 p }{(2\pi)^4}
\Big\langle
N_u^2 \delta_u \w^+
N_d^2 \delta_d \w^-
N_u  {\w}^+
N_d {\w}^-
+
N_u^2 \delta_u \w^+
N_d \w^-
N_u  \w^+
N_d^2 \delta_d \w^-
\nonumber \\ && ~~
+
N_u  \w^+
N_d^2 \delta_d \w^-
N_u^2 \delta_u \w^+
N_d \w^-
+
N_u  \w^+
N_d \w^-
N_u^2 \delta_u \w^+
N_d^2 \delta_d \w^-
\Big\rangle
\nonumber \\ && ~~
+\text{CP invariant terms}
\nonumber  \\
&=&
-\frac{1}{2}
\int\frac{d^4 x d^4 p }{(2\pi)^4}
\tr(N_u N_d N_u^2 N_d^2)
\nonumber \\ && \times
\Big\langle
 \delta_u \w^+ \delta_d \w^- \w^+ \w^-
-
\delta_u \w^+ \w^- \w^+ \delta_d \w^-
- \w^+  \delta_d \w^- \delta_u \w^+ \w^-
\nonumber \\ && ~~
+ \w^+ \w^- \delta_u \w^+ \delta_d \w^- 
\Big\rangle
+\text{CP i.t.}
\label{eq:9.8}
\end{eqnarray}
In the second equality we have rearranged the factors $N_u$, $N_u^2$,
$N_d$, $N_d^2$, using that the CP violating part of the momentum
integral, ${\hat I}_{r_1,t_1,r_2,t_2}^k$, is antisymmetric under
exchange of $r_1$,$r_2$ or $t_1$,$t_2$.

The integrand in (\ref{eq:9.8}) contains derivatives (inside
$\delta_{u,d}$) which are not derivating anything yet. As explained in
section \ref{subsec:2.E}, in general one proceeds by moving the
derivatives to the right, and at the end the momentum integral kills
these ``free'' derivatives. In the present case this turns out not to
be necessary. Instead, we can introduce the combinations
\begin{equation}
(\delta \w)^+ = \delta_u \w^+ - \w^+ \delta_d
,
\qquad
(\delta \w)^- = \delta_d \w^- - \w^- \delta_u
,
\end{equation}
in such a way that
\begin{eqnarray}
\llangle \log K\rrangle_{4+2}
&=&
-\frac{1}{2}
\int\frac{d^4 x d^4 p }{(2\pi)^4}
\tr(N_u N_d N_u^2 N_d^2)
\llangle
 (\delta \w)^+ \w^- (\delta \w)^+ \w^-
-
\w^+ (\delta \w)^- \w^+ (\delta \w)^-
\rrangle
\nonumber \\ &&
+\text{~CP i.t.}
\end{eqnarray}
and the integrand no longer contains any ``free'' derivative.

The CP violating part of this result is proportional to 
${\hat I}_{1,1,2,2}^3$ and develops a factor $(v/\phi(x))^2$ upon
restoration of the Higgs.

There is a simple observation that can already be made at the
present stage. Namely, {\em there is no CP violating contribution to
  the abnormal parity sector} from terms of the type $4+2$ without
Higgs field derivatives. As discussed in section \ref{sec:4}, in the
abnormal parity sector the operator multiplying the momentum integral
must be {\em hermitian} to have a CP violating contribution, however,
\begin{eqnarray}
{\cal O}= 
\tr\left[\gamma_5
\left( 
 (\delta \w)^+ \w^- (\delta \w)^+ \w^-
-
\w^+ (\delta \w)^- \w^+ (\delta \w)^-
\right)
\right]
\label{eq:9.11}
\end{eqnarray}
is purely imaginary. To verify this, we take the complex conjugate of
everything inside ${\cal O}$. In Euclidean space four-dimensional
space $\gamma_\mu$ and $\gamma_5$ are related to their complex
conjugates by a common similarity transformation,
\begin{equation}
\gamma_\mu^* = C_c^{-1}\gamma_\mu C_c,
  \quad
\gamma_5^* = C_c^{-1}\gamma_5 C_c.
\end{equation}
Also, one verifies that
\begin{equation}
(\w^\pm)^*  = C_c^{-1}\w^\mp C_c ,
\quad
((\delta \w)^\pm)^*  = -C_c^{-1} (\delta \w)^\mp C_c .
\end{equation}
Therefore
\begin{equation}
{\cal O}^*= -{\cal O} ,
\end{equation}
as advertised.

In the absence of $F^{u,d}_{\mu\nu}$ and of complex quark mass
matrices, complex conjugation becomes equivalent to
$ud$-parity.\footnote{Under complex conjugation $W^\pm_\mu\to
  -W^\mp_\mu$, $Z_\mu\to -Z_\mu$ (in Euclidean space) while under
  $ud$-parity $W^\pm_\mu\to W^\mp_\mu$, $Z_\mu\to -Z_\mu$.} So ${\cal
  O}$ is imaginary (in the normal and in the abnormal parity sectors)
because it is odd under $ud$-parity. In turn this was obvious without
further calculation once the momentum integral ${\hat I}_{1,1,2,2}^3$
was obtained in (\ref{eq:9.8}). This is because the latter is odd
under $ud$-parity and the full effective action is even (cf. section
\ref{subsec:3.C}).

Remarkably, the operator $\tr[\gamma_5 (\delta \w)^+ \w^- (\delta
  \w)^+ \w^-]$ (and hence its complex conjugate) vanishes by itself
  after taking an angular average and the Dirac trace. We have not
  found a simple explanation for this.

  The operator ${\cal O}$ in the normal parity sector (i.e., as in
  (\ref{eq:9.11}) without $\gamma_5$) is also odd under $ud$-parity
  and so also purely imaginary. Therefore the operation of taking the
  real part indicated in (\ref{eq:8.6}) is redundant. The normal
  parity contribution is not vanishing. The result so obtained is part
  of the general result which we present in the next section.

\section{CP violating terms to six derivatives}
\label{sec:10}

In this section we present the full result for the CP violating terms
of the effective action to six derivatives. This includes all
relevant fields in the SM, and derivatives of the Higgs field.

We have used the method of symbols and repeated the calculation using
the method of covariant symbols as a check, to obtain precisely the
same result from both calculations. In the latter case we use the
covariant derivatives $D_{u,\mu}$ and $D_{d,\mu}$ to carry out the
construction indicated in (\ref{eq:2.27}). This full result is also
consistent with the independent computation made in the previous
section.

From the calculation we obtain the remarkable result that $\llangle
\log K\rrangle_{4+2}$ vanishes identically in the abnormal parity
sector for all terms that could have a contribution to CP violation.
In fact, $\Gamma^-$ vanishes for all the terms that we have computed,
whether CP violating or not. (Of course we have not studied most CP
invariant terms with six or less derivatives, as they are not required
for our purposes.)  We have not found a compelling reason for this, so
most likely, the vanishing found is just a low order accidental
symmetry. The existence of CP violating terms in the abnormal parity
sector of the SM with eight derivatives or more is not excluded. These
would be the leading P violating contributions, relevant to the
electric dipole moment problem.

Another unexpected result is that $\llangle \log K\rrangle_{4+2}$ is
purely real in the normal parity sector, for terms that
contribute to CP violation. (And so, taking the real part indicated in
(\ref{eq:8.6}) becomes redundant.)  In the calculation this follows
from the fact that only the momentum integral $\hat{I}^3_{1,1,2,2}$
appears. This integral is odd under $ud$-parity and hence the
accompanying operator must be odd too and hence imaginary. Once again
it not obvious to us whether this feature will be maintained at higher
orders in the derivative expansion, coming from an exact selection
rule in the SM, or is just an accidental symmetry.

The result, in Minkowski space and in the unitary gauge, reads
\begin{equation}
\Gamma_{\text{SM}} =
-\frac{N_c}{2}
i J\,G_F\,\kCP
\int d^4x \, \left(\frac{v}{\phi} \right)^2
\big(
{\cal O}_0
+
{\cal O}_1
+
{\cal O}_2
\big)
+{\cal O}(D^8)
+\text{CP invariant terms}
\,.
\label{eq:10.1}
\end{equation}
Here $N_c=3$ is the number of colors, $J$ the Jarlskog invariant,
$G_F$ the Fermi constant and $\kCP=3.1\times 10^2$ is the
dimensionless parameter of section \ref{sec:5}. The operators ${\cal
  O}_i$, $i=0,1,2$, have dimension six and are all purely imaginary:

 \begin{eqnarray}
 {\cal O}_0
&=&
 \frac{2}{3} W^+_{\mu }W^-_{\mu \nu }W^+_{\alpha}W^-_{\nu \alpha }
 -\frac{2}{3} W^+_{\mu }W^-_{\mu\nu }W^+_{\alpha }W^-_{\alpha \nu }
 \nonumber\\ && 
 +\frac{4}{3} W^+_{\mu}W^-_{\nu \mu }W^+_{\nu }W^-_{\alpha \alpha}
-2 W^+_{\mu }W^-_{\nu \mu }W^+_{\alpha }W^-_{\nu\alpha }
 \nonumber\\ && 
 +\frac{2}{3} W^+_{\mu }W^-_{\nu \mu }W^+_{\alpha}W^-_{\alpha \nu }
 -\frac{1}{3} W^+_{\mu }W^-_{\nu\nu }W^+_{\mu }W^-_{\alpha \alpha }
 \nonumber\\ && 
 +\frac{5}{3} W^+_{\mu}W^-_{\nu \alpha }W^+_{\mu }W^-_{\nu \alpha}
 -\frac{1}{3} W^+_{\mu }W^-_{\nu \alpha }W^+_{\mu}W^-_{\alpha \nu }
-
\text{c.c.}
 \nonumber\\
{\cal O}_1
&=&
\frac{8}{3} (\varphi_{\mu }+iZ_\mu)
 \nonumber\\ && 
\times\Big(
 W^+_{\nu }W^-_{\mu}W^+_{\nu }W^-_{\alpha \alpha }
 - W^+_{\nu}W^-_{\nu }W^+_{\mu }W^-_{\alpha \alpha }
 \nonumber\\ && 
~~ + W^+_{\nu }W^-_{\alpha }W^+_{\mu}W^-_{\alpha \nu }
 - W^+_{\nu }W^-_{\alpha}W^+_{\nu }W^-_{\mu \alpha }
 \nonumber\\ && 
~~  - W^+_{\nu}W^-_{\alpha }W^+_{\nu }W^-_{\alpha \mu }
 + W^+_{\nu }W^-_{\alpha }W^+_{\alpha}W^-_{\mu \nu }
\Big)
-
\text{c.c.}
 \nonumber\\
{\cal O}_2
&=&
 -\frac{4}{3} (\varphi_{\mu }+iZ_\mu)(\varphi_{\mu }+iZ_\mu)
\Big(
W^+_{\nu }W^-_{\nu }W^+_{\alpha}W^-_{\alpha }
-2 W^+_{\nu}W^-_{\alpha }W^+_{\nu }W^-_{\alpha }
\Big)
 \nonumber\\ && 
 -\frac{4}{3} (\varphi_{\mu }+iZ_\mu)(\varphi_{\nu }+iZ_\nu)
\label{eq:10.4}
\\  && 
\times\Big(
W^+_{\mu }W^-_{\alpha }W^+_{\nu}W^-_{\alpha }
 -2 W^+_{\alpha}W^-_{\mu }W^+_{\alpha }W^-_{\nu }
+2 W^+_{\alpha }W^-_{\nu }W^+_{\mu}W^-_{\alpha }
\Big)
-
\text{c.c.}
\nonumber
\end{eqnarray} 

In these expressions $\text{c.c}$ stands for complex conjugate. Even
if these expressions refer to Minkowski space, the Lorentz indices are
all written as subindices for clarity as no ambiguity may arise.  With
the conventions given above, the pass from Euclidean to Minkowskian
metric amounts to the replacement $Z_\mu \to i Z_\mu$ with no other
change.

As  noted  in  section   \ref{subsec:2.D},  the  effective  action  in
Minkowski  space is  purely real,  at  every order  in the  derivative
expansion and this property is found here.

The fields $W^\pm_{\mu \nu }$ were defined in section \ref{subsec:3.A}
and they are expressed in terms of the $\text{U}_{\text{e.m.}}(1)$
covariant derivative in (\ref{eq:3.17}). On the other hand
$\varphi_\mu$ was defined in (\ref{eq:7.2}) as the logarithmic
derivative of the Higgs field.

As it is readily verified, all the operators ${\cal O}_{0,1,2}$ are
indeed odd under the CP transformation
\begin{eqnarray}
&&
\varphi_\mu(x)\pm i Z_\mu(x) 
\to 
\pi_\mu{}^\nu (\varphi_\nu(\tilde{x})\mp i Z_\nu(\tilde{x}))
,
\nonumber \\
&&
W^\pm_\mu(x) \to -\pi_\mu{}^\nu \, W^\mp_\nu(\tilde{x})
,\qquad
W^\pm_{\mu\nu}(x) \to 
-\pi_\mu{}^\alpha \pi_\nu{}^\beta \, W^\mp_{\alpha\beta}(\tilde{x})
\,.
\end{eqnarray}

Terms including the field strengths $F^{u,d}_{\mu\nu}$ are absent.
This can be understood from the fact that the available operators,
$iF^{u,d}_{\mu\nu}W^+_\mu W^-_\nu W^+_\alpha W^-_\alpha$, are CP even.

On the other hand, the result in (\ref{eq:10.4}) presents some
regularities which for us remain purely ``empirical''.  In ${\cal
  O}_0$ terms coupling $W^+_{\mu\nu}$ to $W^-_{\alpha\beta}$ do not
appear.  It is always possible to change variables from $\varphi_\mu$
and $Z_\mu$ to $\varphi_\mu \pm i Z_\mu$, however, it is not obvious
why, in ${\cal O}_1$, the combinations $\varphi_\mu \pm i Z_\mu$
couple only to $W^\mp_{\alpha\beta}$ and not to $W^\pm_{\alpha\beta}$.
Also it is not clear why in ${\cal O}_2$, the combinations
$\varphi_\mu \pm i Z_\mu$ couple only with themselves and not with
$\varphi_\mu \mp i Z_\mu$.  This latter observation suggests the
speculation that the effective action (or perhaps $\Gamma_-^+$) to
all orders could be of the form $ F[\varphi_\mu + iZ_\mu]-
F[\varphi_\mu - iZ_\mu]$, where the functional $F$ would depend
analytically (holomorphically) on its argument.

Also, the relatively simple dependence of the result on the
combinations $\varphi_\mu \pm i Z_\mu$ suggests the possibility of
reconstructing the full result with Higgs derivatives from that
without $\varphi_\mu$ (by some kind of gauging). In this case the
calculation in section \ref{sec:9} could perhaps be adapted to include
$\varphi_\mu$. We have not tried this in this work.

The fields $\varphi_\mu-i Z_\mu$ and $iW^+_\mu$ follow from projection
of $\nabla_\mu\Phi$ onto $\Phi$ and $\tilde\Phi$, respectively, where
$\nabla_\mu$ represents the full $\text{SU}_L(2)\times \text{U}_Y(1)$
covariant derivative, and $W^+_{\mu\nu}$ can also be written using
$\Phi$ and $\tilde\Phi$ and their covariant derivatives, but the
result is not particularly illuminating.

\section{The coefficient $\kCP$}
\label{sec:12}

In this section we study in some depth the function $\kCP$. Using the
second form in (\ref{eq:6.1}) as well as the identity (\ref{eq:5.9}),
the momentum integral $\hat{I}_{1,1,2,2}^3$ takes the form
\begin{equation}
\hat{I}_{1,1,2,2}^3
= iJ \sum_{i,j,k,l=1}^3 \! \epsilon_{ik}\epsilon_{jl}
\int \frac{d^4p}{(2\pi)^4}(p^2)^3
\frac{1}{(p^2+m_{u,i}^2)}
\frac{1}{(p^2+m_{d,j}^2)}
\frac{1}{(p^2+m_{u,k}^2)^2}
\frac{1}{(p^2+m_{d,l}^2)^2}
.
\end{equation}
Here $m_{u,i}$ denotes the mass of the quark of type $u$ of the $i$-th
generation, and similarly for $m_{d,i}$.

The sum over flavors is easily carried out using the identity
\begin{equation}
\sum_{i,k}^3 \! \epsilon_{ik}
\frac{1}{(p^2+m_{u,i}^2)}
\frac{1}{(p^2+m_{u,k}^2)^2}
=
-\frac{
(m_u^2-m_c^2)(m_c^2-m_t^2)(m_t^2-m_u^2)
}{
(p^2+m_u^2)^2(p^2+m_c^2)^2(p^2+m_t^2)^2
}
,
\end{equation}
and similarly for $d$-type quarks. The integral can then be written as
\begin{eqnarray}
 \hat{I}_{1,1,2,2}^3
&=&
 iJ \,\Delta_m \, I_m
.
\end{eqnarray}
where
\begin{eqnarray}
\Delta_m &=&
(m_u^2-m_c^2)(m_c^2-m_t^2)(m_t^2-m_u^2)(m_d^2-m_s^2)(m_s^2-m_b^2)(m_b^2-m_d^2)
\,,
\\
I_m
&=&
\int \frac{d^4p}{(2\pi)^4}(p^2)^3
\prod_{q=1}^6 \frac{1}
{(p^2+m_q^2)^2}
.
\label{eq:12.3}
\end{eqnarray}

The factor $J\,\Delta_m$ is just the Jarlskog determinant $\Delta$ of
(\ref{eq:1.1}). This is the ``minimal'' factor that must always be
present in the CP odd effective action as dictated by perturbation
theory. This can be seen for instance considering the same integral at
finite temperature which amounts to replace the energy integral by a
fermionic Matsubara sum. In the limit of sufficiently high
temperature all quark masses can be treated as perturbations. So the
known perturbative result applies and it starts at order twelve with
$J\,\Delta_m$. As a consequence the integral $\hat{I}_{1,1,2,2}^3$ is
consistent with the well known fact that no CP breaking can take place
if two up-type or two down-type quark have the same finite mass. (The
case of degeneracy with vanishing mass requires a separate study since
$I_m$ could present infrared divergences.)

In $I_m$ the product is over the six quark flavors, regardless of its
$u$ or $d$ type. So the coefficient $\kCP$ introduced in section
\ref{sec:5} has the same symmetry as $\Delta_m$ under exchange of
quark flavors, the factor $I_m$ being completely symmetric, and
actually positive definite.

The momentum integrals of the type $I_m$, typical of zero momentum
insertions in a Feynman graph, can be computed using the relation
\cite{Salcedo:2008bs}
\begin{equation}
\int \frac{d^dp}{(2\pi)^d}(p^2)^k\prod_{j=1}^n\frac{1}{(p^2+\sm_j^2)^{r_j}}
=
\frac{(-1)^{k+d/2-1+\sum_j r_j}}{(4\pi)^{d/2}\Gamma(d/2)}
{\cal I}^{k+d/2-1}_{r_1,\ldots,r_n}
, 
\label{eq:12.1}
\qquad
k+d/2=1,2,3,\ldots
\end{equation}
where
\begin{equation}
  {\cal I}^\alpha_{r_1,\ldots,r_n}(m_1,\ldots,m_n) = 
\oint \frac{dz}{2\pi i} \,z^\alpha \log(z) 
\prod_{j=1}^n\frac{1}{(z-\sm_j^2)^{r_j}} \,,
\label{eq:12.2}
\end{equation}
and the integration is along a positive closed simple contour
enclosing the poles at $\sm_j^2$ but excluding $z=0$, and the cut is
on the real negative axis. The identity (\ref{eq:12.1}) assumes
positive $m_j^2$ and holds whenever the left-hand side is ultraviolet
and infrared finite. If it is not, the right-hand side gives the
finite part.\footnote{Note that when $\alpha$ is a non negative
  integer, the contour integral scales as $1+\alpha-\sum_j r_j$ but in
  the presence of infrared divergencies an anomalous scale term
  develops from the logarithm .}

Using (\ref{eq:12.1}), and comparing with the definition of the loop
function $\kCP$ in (\ref{eq:5.11}) yields
\begin{equation}
\kCP= 2^{3/2}\Delta_y \,I_y
,
\end{equation}
where
\begin{eqnarray}
  \Delta_y &=&
  (y_u^2-y_c^2)(y_c^2-y_t^2)(y_t^2-y_u^2)
  (y_d^2-y_s^2)(y_s^2-y_b^2)(y_b^2-y_d^2)
=(\sqrt{2}/v)^{12}\,\Delta_m
,
\nonumber \\
I_y &=&
\frac{1}{(4\pi)^2}\, {\cal I}^4_{2,2,2,2,2,2}(y_u,y_c,y_t,y_d,y_s,y_b)
=(v/\sqrt{2})^{14}\,I_m
  \,,
\end{eqnarray}
and the Yukawa couplings $y_q=\sqrt{2}m_q/v$ are used.  The coupling
controlling the strength of the CP violating operators has dimensions
of one over mass squared. The use of the Yukawa coupling amounts to
using $v$ as convenient mass scale to measure this coupling.

The contour integrals (\ref{eq:12.2}) are readily computed by
residues.\footnote{
$${\cal I}^\alpha_{r_1,\ldots,r_n}(m_1,\ldots,m_n) = 
\sum_{i=1}^n
\frac{1}{(r_i-1)!}\frac{d^{r_i-1}}{d(\sm_i^2)^{r_i-1}}
\frac{(\sm_i^2)^\alpha \log(\sm_i^2) }
{\prod_{j=1,j\not= i}^n(\sm_i^2-\sm_j^2)^{r_j}}
\,.
$$} This produces the explicit expression
\begin{eqnarray}
I_y
&=&
\frac{1}{(4\pi)^2}\,
\sum_i
\Bigg[
\frac{y_i^6(1+4\log y_i^2)}{\prod^\prime_j (y_i^2-y_j^2)^2}
-\frac{
y_i^8 \log y_i^2
}{
\prod^\prime_j (y_i^2-y_j^2)^3
}
\Big(
10 y_i^8
-8 y_i^6 \sum_j \!{}^\prime \, y_j^2 
\nonumber \\ && 
+ 6y_i^4 \sum_{j<k}\!{}^\prime \, y_j^2 y_k^2
- 4 y_i^2 \sum_{j<k<l}\!\!\!{}^\prime \,\, y_j^2 y_k^2 y_l^2
+ 2 \sum_{j<k<l<r}\!\!\!\!\!\!{}^\prime \,\,\, y_j^2 y_k^2 y_l^2 y_r^2
\Big)
\Bigg]
.
\end{eqnarray}
In this expression the indices $i,j,k,l,r$ run over the six quark
flavors.  The prime in a sum or product indicates to omit the term $i$
in that sum or product. Although the explicit expression of the
integral $I_y$ looks divergent as two masses become degenerated this
is not so, as is obvious from the integral itself: in the coincidence
limit the integral (\ref{eq:12.3}) is perfectly regular. Also, $I_y$ yields an
homogeneous function of the $y$ of degree $-14$.  The inhomogeneous
scale variation from the logarithms cancels after adding the six
terms. (This implies that one can use one of the Yukawa couplings,
$y_q$, as overall scale, $y_i\to y_i/y_q$ to remove one of the six
logarithms in the expression, at the price of a less symmetric
formula.) The formula can be written in many different ways but none
is expected to give a simple expression. The simplest and most
transparent form is perhaps its very definition as a momentum integral,
(\ref{eq:12.3}).

Each of the factors $\Delta_y$, $I_y$ in $\kCP$ gives quite disparate
numbers, namely, $\Delta_y= 6.0\times 10^{-18}$ and $I_y= 1.8\times
10^{19}$.  This is because they are homogeneous functions of very
large degree (high mass dimension) and so they change wildly under
even moderate changes in the scale. For instance, in units of the
bottom quark mass $\Delta_m=1.5\times 10^2$ and $I_m=4.1\times
10^{-4}$. Their product has degree $-2$ and so the number is much less
dependent on the mass scale.

Therefore the huge cancellation between factors is partially trivial,
and as noted in \cite{Smit:2004kh} the factor $\Delta_y$ should not be
used as a rough estimate of the CP violating component. What is not
trivial is the detailed role played by the light quarks. The value of
$\kCP$ is enhanced by the small mass of the quarks $u$, $d$ and $s$.
This can be seen by changing artificially their mass, $m_u\to \lambda
m_u$, $m_d\to \lambda m_d$, $m_s\to \lambda m_s$. As $\lambda$ moves
in the range from $1$ to $25$ one finds that $I_y$ decreases
monotonously by a factor $5000$. This rather large factor corresponds
to an effective dimension which runs from about $-2$ at $\lambda=1$ to
about $-2.5$ at the highest value of $\lambda$. The value $-2$
corresponds to the scale dimension at the infrared divergent point
$\lambda=0$. On the other hand the factor $\Delta_y$ increases by a
factor about 400. Beyond this range $\Delta_y$ reaches a maximum and
starts to fall heading for the zero at $\lambda=m_b/m_s\sim 40$, and
then starts to grow again with opposite sign. The net effect is to
find a quenching of $\kCP$ as one departs from the SU$(3)$ chiral
point. Nevertheless, it should be noted that $\kCP$ is a rather
complicated function of the quark masses, without well defined sign,
so any interpretation should be taken with some caution as the result
might look different depending on how the masses are moved in detail.

The coefficient $\kCP$ is infrared finite as any two quark masses go
to zero while the other quarks remain massive.  This allows us to
consider the limiting case $m_u=m_d=0$.  (The coefficient vanishes
trivially if the two massless quarks are $u$-like or $d$-like, due to
the antisymmetry of the $\Delta_y$ factor.)  This gives for the $I_y$
factor
\begin{eqnarray}
I_y &=&
\frac{1}{(4\pi)^2}\,
{\cal I}^0_{2,2,2,2}(y_c,y_t,y_s,y_b)
 \\ 
&=&
\frac{1}{(4\pi)^2}\,
\sum_i
\Bigg[
\frac{1}{y_i^2}\frac{1}{\prod^\prime_j (y_i^2-y_j^2)^2}
+\frac{\log y_i^2}{\prod^\prime_j (y_i^2-y_j^2)^3 }
\Big(
-6 y_i^4
+4 y_i^2 \sum_j \!{}^\prime \, y_j^2 
- 2 \sum_{j<k}\!{}^\prime \, y_j^2 y_k^2
\Big)
\Bigg]
.
\nonumber
\end{eqnarray}
Here the indices $i,j,k$ run over the four remaining flavors $s,c,b,t$.

The approximation $m_u,m_d\to 0$ does not change much the value of the
coefficient $\kCP$, overestimating it by a $3\%$ as compared to the
exact value.  Since the mass of the quark $s$ is also rather small one
can consider the further limit $m_s\to 0$. The limit of vanishing
$m_u,m_d,m_s$ is affected by infrared divergencies and this manifests
in the fact that while $\kCP$ remains finite, the value depends on how
this limit is taken.  Explicitly, the dominant term as $m_u,m_d,m_s$
become small is
\begin{eqnarray}
\Delta_y I_y &=&
\frac{1}{(4\pi)^2}\,
\Bigg[
\frac{
-y_d^4 y_s^4 (y_d^2 + y_s^2)
+2 y_u^2 y_d^2 y_s^2 (y_d^4 + y_s^4)
-y_u^4(y_d^6 + y_s^6) 
- y_u^6(y_d^2 - y_s^2)^2
}{(y_d^2 - y_s^2) (y_u^2-y_d^2)^2 (y_u^2-y_s^2)^2}
\nonumber \\ &&
- 2 y_d^6  \log (y_d^2/y_u^2)  
\frac{ 2 y_u^2 y_s^2 - y_u^2 y_d^2  - y_d^2 y_s^2
}
{(y_d^2 - y_s^2 )^2 ( y_d^2 -y_u^2 )^3}
\label{eq:12.4}
\\ &&
+ 2 y_s^6  \log (y_s^2/y_u^2)  
\frac{ 2 y_u^2 y_d^2 - y_u^2 y_s^2  - y_d^2 y_s^2
}
{(y_d^2 - y_s^2 )^2 ( y_s^2 -y_u^2 )^3}
\Bigg]
\left(
\frac{1}{y_c^2}-\frac{1}{y_t^2}
\right)
+O(y^2_{{\text{\rm light quark}}})
.
\nonumber 
\end{eqnarray}
The terms retained are those which are homogenous of degree zero in
$y_u,y_d,y_s$, while the reminder has degree 2.  This function is
antisymmetric under the exchange of $d$ and $s$. A quite remarkable
fact is that, to this order, the dependence on $y_c$ and $y_t$
factorizes and moreover $y_b$ does not appear in the expression.

The above function is finite but far from continuous in the massless
limit. The physical situation is $m_u,m_d\ll m_s$
so a sensible limit to consider corresponds to taking first $m_u,m_d\to
0$ and later $m_s\to 0$. In this approximation
\begin{equation}
\Delta_y I_y = 
\frac{1}{(4\pi)^2}
\left(
\frac{1}{y_c^2}-\frac{1}{y_t^2}
\right)
\,.
\end{equation}
Equivalently, in this SU(3) chiral limit
\begin{equation}
G_F\, \kCP
=
\frac{1}{(4\pi)^2}\,
\left(\frac{1}{m_c^2}-\frac{1}{m_t^2}\right)
\,.
\end{equation}
The full result corresponds rather to $0.92/((4\pi)^2 m_c^2)$ so this
approximation overestimates it by an $8\%$. Despite the simplicity of
this approximation the massless SU(3) chiral limit is highly non
trivial as can be seen from (\ref{eq:12.4}), and in fact even the sign
is not obvious from the expression.

\section{Conclusions}
\label{sec:11}

We have studied the CP breaking terms in the effective action of the
Standard Model obtained by integration of quarks and leptons,
including operators up to dimension six. The result of the calculation
is summarized in eqs. (\ref{eq:10.1}) and (\ref{eq:10.4}).

One of the main results is that such CP breaking operators appear with
a sizable coupling, of the order of $5\times 10^2$ times the Jarlskog
invariant times the Fermi constant. This is much larger than predicted
from Jarlskog determinant considerations based on perturbation theory
and is fully in agreement with the expectations first put forward by
Smit in \cite{Smit:2004kh}.

Remarkably the non vanishing CP violating contributions come from the
normal parity sector. This is is somewhat unexpected as it is usually
taken for granted that the presence of the Levi-Civita pseudo-tensor
is needed to have CP breaking. Also noteworthy is our finding that, to
the order studied, all abnormal parity terms vanish. This is in
conflict with the result presented in \cite{Hernandez:2008db} where a
non vanishing abnormal parity contribution is derived. Our result
implies that the first CP odd and P odd contribution, relevant to
electric dipole moments, requires at least the eighth order in the
derivative expansion.

The full result presents interesting regularities (including the just
mentioned vanishing of terms which are simultaneously CP odd and
parity odd) which may follow from the structure of the SM or
may be accidental symmetries surviving only at lowest orders in the
derivative expansion.

The calculation presented here applies to zero temperature. At the
order studied in this work, the coupling is controlled by the integral
$\hat{I}^3_{1,1,2,2}$. The same integral can be considered at finite
temperature replacing the energy integral in the loop by a fermionic
Matsubara sum. At high enough temperatures such integral becomes
proportional to the Jarlskog determinant times the accompanying power
$1/T^{14}$, just by dimensional counting. For temperatures comparable
to the top quark mass, relevant for baryogenesis, this mass can no
longer be treated as a perturbation.  Nevertheless, the integral
becomes equal to the Jarlskog determinant times an order of unity
function of $m_t/T$ and the required power $1/T^{14}$, so the
numerical estimate is similarly small. At higher order in the
derivative expansion other momentum integrals will be generated but
the analysis is expected to be similar.  On the other hand, at finite
temperature new operators can be produced even at the same order
studied here. Due the to breaking of Lorentz invariance down to
rotational invariance \cite{Megias:2002vr} new combinations may arise,
e.g., with abnormal parity.

The effective action obtained here can not be applied directly to CP
violating hadronic processes since the quarks have been integrated
out.  In this view it would be of interest to repeat the
calculation but inserting external hadronic currents in the quark
sector. This would allow to account for CP violating contributions to
meson decay amplitudes. CP violation requires to switch between the
$u$ and $d$ spaces and in the present calculation this can only be
achieved by the action of $W^\pm$ which are the only charged particles
left after integration of fermions. The situation may change in the
presence of hadronic insertions carrying electric charge, and
presumably lower order operators would be allowed replacing some
charged gauge bosons by hadronic currents.

Related to the above, we have seen that the coupling $\kCP$
controlling the strength of the CP violating operators is a
complicated function of the quark masses and is highly non continuous
in the relevant limit of taking the light quarks to be massless. This
is due to the presence of infrared divergences in the chiral limit.
This suggests that the standard chiral perturbation theory corrections
induced by the pseudo-scalar Goldstone bosons could introduce sizable
modifications to the result. Formally, we have included QCD in our result,
although no gluonic correction have been needed to the order
considered. Of course, the calculation can be organized under
different schemes, for instance counting independently derivatives
coming from electroweak fields and QCD fields, being the latter either
fundamental or hadronic. A point to note in this regard is that in the
presence of hadrons abnormal parity is no longer equivalent to parity
violating, as the pseudo-scalars, for instance, have abnormal parity.

Likewise it would be of interest to consider the effective action in
the sense of the Legendre transform (after integration of fermions),
i.e., adding one particle irreducible graphs. This could produce lower
dimensional CP breaking operators, such as the operator
$\tr(F_{\mu\nu}\tilde{F}^{\mu\nu})$ mentioned in the Introduction.
This can be so by integration of some of the external lines in the
loop (hence making these lines internal) either in a single fermion
loop having CP violation or by coupling more than one loop, one of
them carrying CP violation.

As a final comment we note that leptons have not been included as no
CP breaking takes place in that sector when the neutrino masses
vanish.  However, our calculation applies to Dirac neutrinos with non
zero masses, and in this case the leptons produce a CP violating
contribution completely similar to that obtained for quarks, replacing
the Jarlskog invariant of the CKM matrix with that of the
Maki-Nakagawa-Sakata matrix \cite{Maki:1962mu}, and using lepton
masses to produce $\kCP^{\text{leptons}}$. This function will be
identical to that discussed in section \ref{sec:12} for quarks.
Similarly as for the case of quarks, the coefficient
$\kCP^{\text{leptons}}$ is not at all continuous as some of the
leptons become massless. This implies that the limit will depend
crucially on how the small neutrino masses compare with each other.

\acknowledgments
This work is supported in part by funds provided by the Spanish DGI
and FEDER funds with grant  FIS2008-01143, Junta de Andaluc{\'\i}a
grants FQM225, FQM481 and P06-FQM-01735 and EU Integrated
Infrastructure Initiative Hadron Physics Project contract
RII3-CT-2004-506078.


\providecommand{\href}[2]{#2}\begingroup\raggedright\endgroup

\end{document}